\documentclass[12pt,letterpaper]{article}
\usepackage[a4paper, total={7in, 10in}]{geometry}

\usepackage{graphicx}
\usepackage{helvet}
\usepackage{authblk}
\usepackage{hyperref}
\usepackage{amsmath} 
\usepackage{amssymb} 
\usepackage{orcidlink} 
\usepackage[super,comma,sort&compress]  
   {natbib}
\usepackage[title]{appendix}%
\usepackage{float}

\usepackage[utf8]{inputenc} % allow utf-8 input
\usepackage[T1]{fontenc}    % use 8-bit T1 fonts
\usepackage{url}            % simple URL typesetting
\usepackage{booktabs}       % professional-quality tables
\usepackage{amsfonts}       % blackboard math symbols
\usepackage{nicefrac}       % compact symbols for 1/2, etc.
\usepackage{microtype}      % microtypography
\usepackage{xcolor}         % colors
\usepackage{multirow}
\usepackage{threeparttable}
\usepackage{amsmath}
\usepackage{subcaption}
\usepackage{makecell}
\usepackage{placeins}
\usepackage[utf8]{inputenc} % 支持 UTF-8 编码
\usepackage{booktabs}      % 用于生成专业的三线表
\usepackage{siunitx}       % 用于按小数点对齐数字
\usepackage{caption}       % 用于表格标题
\usepackage{tabularx}
\usepackage{listings}    % 引入 listings 宏包
\usepackage{pifont}
\usepackage{array}
\usepackage{makecell}
\usepackage{caption}
\usepackage{pdfpages}
\usepackage{placeins} 
\usepackage{rotating}
\usepackage{booktabs,multirow,adjustbox,makecell}
\theadset{\bfseries}

\makeatletter
\renewcommand{\maketitle}{\bgroup\setlength{\parindent}{0pt}
\begin{flushleft}
  \textbf{\@title}
  
  \@author
\end{flushleft}\egroup}
\makeatother

\title{ECG-SDI: Single-lead Electrocardiogram Enables Continuous Sleep Depth Index Profiling and Scalable Digital Biomarkers for Sleep Health
}
\date{}

\author[1,2,\#,\orcidlink{0009-0008-6988-2183}]{Donglin Xie}
\author[3,\#]{Qingshuo Zhao}
\author[4,\#]{Songchi Zhou}
\author[5]{Haoyu Wang}
\author[6]{Jingyu Wang}
\author[3,7]{Yuxi Zhou}
\author[8,*]{Thomas Penzel}
\author[1,2,9,*,\orcidlink{0000-0001-7521-5127}]{Shenda Hong}

\affil[1]{Institute of Medical Technology, Peking University Health Science Center, Beijing, China}
\affil[2]{National Institute of Health Data Science, Peking University, Beijing, China}
\affil[3]{Department of Computer Science, Tianjin University of Technology, Tianjin, China}
\affil[4]{Department of Statistics and Data Science, Tsinghua University, Beijing, China}
\affil[5]{University of Chinese Academy of Sciences, China}
\affil[6]{Department of Respiratory and Critical Care Medicine, Binzhou Medical University Hospital, Binzhou, China}
\affil[7]{Department of Respiratory, The Second Hospital of Tianjin Medical University, Tianjin, China}
\affil[8]{Interdisciplinary Center of Sleep Medicine, Charité - Universitätsmedizin Berlin, Berlin, Germany}
\affil[9]{Institute for Artificial Intelligence, Peking University, Beijing, China}

\affil[$\#$]{These authors contributed equally}
\affil[*]{Correspondence: hongshenda@pku.edu.cn, thomas.penzel@charite.de}

\begin{document}

\maketitle

% 思路

\section*{ABSTRACT}

Traditional sleep staging provides a discrete and coarse-grained representation of sleep, limiting both the characterization of within-stage depth variation and the scalability of continuous sleep phenotyping to lightweight and home-based settings. Here we propose ECG-SDI, a single-lead Electrocardiogram (ECG) framework for continuous sleep-depth profiling and digital biomarker extraction. Using 12,392 overnight PSG recordings across multiple cohorts, we trained and validated a two-stage deep learning model to learn a continuous sleep depth index ranging from 0 to 1 under supervision from conventional discrete sleep-stage labels. In the pooled internal test set, ECG-SDI achieved a correlation of 0.828 with ordered sleep stages excluding REM and a REM classification AUROC of 0.954, and demonstrated a robust performance in an external CFS cohort with a correlation of 0.848 and an AUROC of 0.977. ECG-SDI preserved whole-night sleep organization and revealed intra-stage physiological heterogeneity that was not apparent from conventional sleep stages alone. Biomarker-based phenotyping identified a disturbed sleep subtype associated with poorer subjective sleep quality and higher all-cause mortality in the SHHS (adjusted hazard ratio, 1.20; 95\% confidence interval, 1.05–1.38). In 507 participants with paired recordings before and after continuous positive airway pressure (CPAP), normalized sleep-depth area increased by 6.1\% and shallow-sleep burden below 0.3 decreased by 16.0\%. Exploratory analyses linked reductions in approximate entropy and depth variability to reduced oxygen desaturation and increased sleep efficiency, respectively (Spearman $\rho = 0.24$ and 0.12); both associations remained positive after covariate adjustment. Mean ECG-SDI within N2 increased by 0.017 (95\% confidence interval, 0.004–0.030). A wearable proof-of-concept illustrated the acquisition-to-report workflow. These findings extend continuous sleep-depth phenotyping to single-lead electrocardiography and provide evidence of responsiveness to CPAP-associated changes, supporting prospective evaluation for repeated home-based assessment.

\section*{KEYWORDS}
Single-lead ECG, Continuous sleep depth, Digital biomarkers, CPAP response

%%%  Include up to 10 keywords, separated by commas. 
%%%  Keywords entered in EM are not carried over; only 
%%%  keywords included in the main text will be used 
%%%  in the final article metadata. Please note that 
%%%  for some journals, keywords are chosen by the editors.

\section*{Introduction}
Sleep health is a fundamental pillar of cardiovascular, metabolic, cognitive, and neuropsychiatric homeostasis. REM sleep contributes to memory processing, while persistent sleep disturbance is closely linked to obesity, depression, neurodegenerative disease, and a broader spectrum of objectively measured disease phenotypes\cite{boyce2017rem,baranwal2023sleep,ogilvie2017epidemiology,riemann2020sleep,hu2020rem,pagano2018rem,wang2025phenome}. At the same time, sleep fragmentation, obstructive sleep apnea (OSA), and poor subjective sleep quality impose substantial daytime functional, cognitive, and emotional burden and contribute to long-term adverse health outcomes\cite{martin1996effect,benkirane2022impact,boon2023effect}. Polysomnography (PSG) remains the clinical gold standard for sleep assessment, yet its acquisition and scoring are resource-intensive, and manual staging still shows inter-scorer variability\cite{fallmann2019computational,magalang2013agreement,younes2016staging}. More fundamentally, conventional sleep staging based on fixed 30-s epochs is a coarse-grained representation that does not adequately capture progressive within-stage variation in sleep depth, nor does it directly quantify sleep fragmentation and recovery dynamics\cite{bonnet1997heart}.

Against this background, digital biomarkers and continuous sleep depth offer a compelling alternative framework for sleep characterization. A biomarker is an objectively measurable characteristic that reflects physiological processes, pathological processes, or responses to intervention; when such characteristics are enabled by digital health technologies for repeated, remote, and continuous acquisition, they gain the practical value of digital biomarkers\cite{mesko2017digital,senbekov2020recent,coravos2019developing,vasudevan2022digital}. Sleep is particularly well suited to this paradigm because nocturnal physiological activity is relatively stable and tightly linked to a broad range of health outcomes\cite{wang2025phenome}. Along the axis of continuous sleep depth, studies of the Odds Ratio Product (ORP), reproducibility of Electroencephalogram (EEG) sleep biomarkers, calls for higher-resolution hypnograms, and more recent data-driven ordinal sleep depth measures all suggest that continuous representations align more naturally with arousal propensity and sleep instability than discrete stage labels, and may be especially useful for tracking treatment response\cite{younes2015odds,penner2019odds,younes2021characteristics,gratton2025toward,meulenbrugge2026ordinal}. Our prior PSG-based SDI study further demonstrated that continuous sleep depth and its derived time-series features can yield new digital biomarkers for sleep health\cite{zhou2025continuous}.

However, if continuous sleep-depth phenotyping still depends on EEG or full PSG, its scalability in home-based, lightweight, and real-world settings remains limited. Single-lead ECG is one of the most widely available physiological signals, is well suited to long-term wear, and is naturally compatible with home deployment; meanwhile, portable and wearable devices have achieved increasingly practical ECG/HRV acquisition, and home sleep monitoring technologies are rapidly maturing\cite{dobbs2019accuracy,ozkan2019portable,yoon2023technologies,mathunjwa2025comprehensive}. From a physiological perspective, stage transitions, REM/NREM reorganization, and arousal are all accompanied by changes in autonomic and cardiovascular regulation, making ECG and HRV intrinsically informative about sleep\cite{bonnet1997heart,da2021heart}. Prior work has already established the feasibility of reduced-sensor sleep analysis: from deep-learning sleep staging with single-channel EEG to heart-rate/HRV-based staging, cardiorespiratory sleep detection, pulse-oximetry-derived sleep-wake classification, EEG+ECG multimodal staging, and single-lead ECG for sleep apnea detection\cite{supratak2017deepsleepnet,fonseca2016cardiorespiratory,radha2019sleep,casal2019sleep,lyu2023novel,akter2026dream}. In parallel, rapid progress in AI-enabled ECG phenotyping suggests that limited-lead ECG can encode clinically meaningful information far beyond conventional heart-rate summaries\cite{zhu2026artificial}. Nevertheless, most existing ECG-based studies still focus on discrete sleep staging or event detection rather than continuous sleep depth itself, and the reliability of wearable-based sleep staging remains imperfect\cite{birrer2024evaluating}.

\begin{figure}
    \centering
    \includegraphics[width=\textwidth]{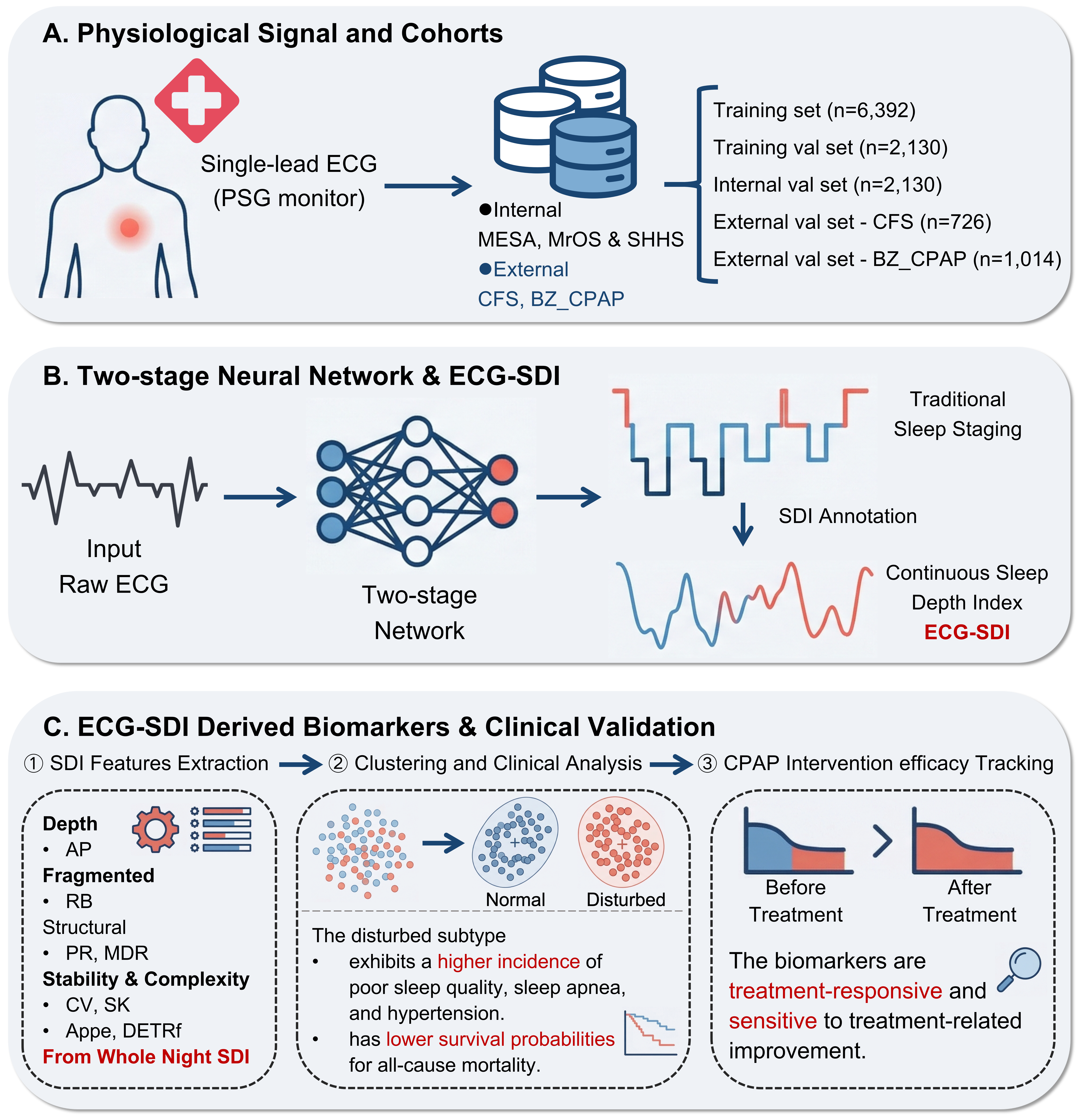}
    \caption{\textbf{Overall study framework and analytical pipeline.} \textbf{a,} Single-lead ECG was extracted from multi-cohort PSG recordings to construct internal datasets for model development (SHHS, MESA, and MrOS) and fully independent external validation datasets (CFS and BZ-CPAP). \textbf{b,} A two-stage neural network was used to learn a continuous sleep depth index (ECG-SDI) from raw ECG under supervision from conventional discrete sleep-stage labels, thereby transforming coarse-grained sleep staging into a continuous representation of sleep depth. \textbf{c,} Digital biomarkers were then derived from whole-night ECG-SDI trajectories and used for sleep-phenotype clustering, clinical risk stratification, and tracking CPAP treatment effects, enabling systematic evaluation of the physiological relevance, clinical utility, and treatment responsiveness of ECG-SDI.}
    \label{fig:overview}
\end{figure}

Accordingly, what is currently missing is not simply another ECG sleep-staging model, but a \emph{single-lead ECG-derived, continuous, interpretable, and treatment-responsive sleep-depth biomarker framework}. Such a framework could move continuous sleep-depth phenotyping beyond the PSG laboratory into more accessible real-world monitoring settings, while also offering a more sensitive means of quantifying treatment-related restoration of sleep continuity---particularly in the context of continuous positive airway pressure (CPAP), which is recommended for adult OSA but still lacks sufficiently refined objective markers for evaluating benefit\cite{patil2019treatment,batool2016impact,penner2019odds}. Fig.~\ref{fig:overview} summarizes the overall study framework, and our contributions are as follows:
\begin{enumerate}
    \item We propose ECG-SDI, a scalable single-lead ECG framework for continuous sleep-depth profiling, and validate it in large public sleep cohorts together with an external clinical CPAP cohort.
    \item We show that ECG-SDI preserves whole-night sleep organization while revealing intra-stage physiological heterogeneity that is not visible from conventional stage labels alone.
    \item We derive multidimensional time-series digital biomarkers from overnight ECG-SDI trajectories and show that their clustering identifies sleep phenotypes associated with adverse sleep-health profiles and increased all-cause mortality.
    \item We demonstrate in the Binzhou Longitudinal CPAP (BZ-CPAP) cohort that these digital biomarkers are sensitive to treatment-related sleep improvement, thereby supporting ECG-SDI not only as a static sleep descriptor but also as a treatment-responsive digital biomarker for longitudinal follow-up and home-based monitoring.
\end{enumerate}

% ---------- Figure 1 ----------

% ---------- Results ----------
% Results fragment for inclusion in the supplied manuscript. Figure 1 remains in the Introduction.
\section*{RESULTS}

\subsection*{Data curation and deep learning model development}

A total of 12,392 overnight PSG recordings were included in the multi-cohort analysis. Of these, 10,652 recordings from SHHS, MESA, and MrOS were divided at the participant level into training, validation, and internal test sets in a 6:2:2 ratio. The independent CFS and BZ-CPAP cohorts contributed 1,740 external recordings and were not used for neural-network training or model selection. CFS contributed 726 recordings, whereas BZ-CPAP comprised 507 participants with recordings obtained before and after CPAP, yielding 1,014 recordings. Detailed demographic information for the various cohorts is provided in the Appendix~\ref{app:info}.

We adapted a two-stage single-lead ECG sleep-staging framework to produce a continuous-valued sleep depth index, ECG-SDI, on a scale from 0 to 1, together with REM-related predictions. The pretrained convolutional and Transformer-based backbone was retained as a frozen feature extractor, and two task-specific output heads were trained using conventional sleep-stage labels. The model generated an ECG-SDI value for each 30-s epoch while incorporating neighboring-epoch context. After model development, the network parameters were fixed for subsequent evaluation of sleep-stage ordering, overnight trajectories, clinical phenotypes, and paired CPAP-associated changes. The architecture and implementation details are provided in Appendix~\ref{app:framework}.

\subsection*{Single-lead ECG robustly recovers continuous sleep depth and REM structure across internal and external cohorts}
As shown in Table~\ref{tab:performance_main}, in the pooled internal test set, ECG-SDI demonstrated a correlation of 0.828 with ordered sleep stages excluding REM and achieved a REM classification AUROC of 0.954. At the cohort level, the corresponding sleep-depth correlations in the internal SHHS, MESA, and MrOS test sets were 0.730, 0.797, and 0.882, with REM classification AUROCs of 0.908, 0.934, and 0.987, respectively. In the fully independent external validation cohorts, performance remained high in CFS, with a correlation of 0.848 and a REM classification AUROC of 0.977. In the BZ-CPAP cohort, the corresponding correlations were 0.456 before CPAP and 0.546 after CPAP, with REM classification AUROCs of 0.803 and 0.830, respectively. Taken together, these findings indicate that ECG-SDI achieved good overall performance, with robust REM discrimination and preserved sleep-depth ordering across heterogeneous cohorts and acquisition settings.

ECG-SDI displayed an expected stage-wise distribution across all four cohorts, which is shown in Fig.~\ref{fig:stage_distribution} of Appendix~\ref{app:box}. Wake epochs were concentrated at the lowest depth range, values increased progressively from N1 to N3, and N3 consistently exhibited the highest sleep depth, whereas REM occupied an intermediate range distinct from both wakefulness and deep NREM sleep. This overall pattern indicates that single-lead ECG alone is sufficient to recover a physiologically meaningful continuum of sleep depth across conventional sleep stages. Notably, the wake distributions in SHHS and MESA were broader than expected and showed an upper tail extending to relatively high depth values. This deviation may partly reflect the cardiovascular burden and related alterations in nocturnal cardiac dynamics in these cohorts; we therefore note this observation here and revisit it in the Discussion.

\begin{table}[t]
\centering
\caption{\textbf{Performance of ECG-SDI for sleep-depth ordering and REM classification in internal and external cohorts.} Corr. SS (excl. REM), correlation between ECG-SDI and ordered sleep stages excluding REM; REM class. AUROC, area under the receiver operating characteristic curve for REM classification. Values are reported as point estimates with corresponding uncertainty intervals.}
\label{tab:performance_main}
\setlength{\tabcolsep}{5pt}
\small
\begin{tabular}{lcc}
\toprule
Cohort & \shortstack[c]{Corr. SS \\ (excl. REM)$\uparrow$} & \shortstack[c]{REM class. \\ AUROC$\uparrow$} \\
\midrule
Pooled internal test                     & 0.828 (0.827--0.828) & 0.954 (0.953--0.954) \\
Internal SHHS                & 0.730 (0.729--0.732) & 0.908 (0.907--0.909) \\
Internal MESA                & 0.797 (0.796--0.798) & 0.934 (0.933--0.935) \\
Internal MrOS                & 0.882 (0.882--0.883) & 0.987 (0.987--0.988) \\
External CFS                 & 0.848 (0.847--0.849) & 0.977 (0.976--0.977) \\
External BZ-CPAP (pre-CPAP)  & 0.456 (0.453--0.459) & 0.803 (0.802--0.805) \\
External BZ-CPAP (post-CPAP) & 0.546 (0.544--0.549) & 0.830 (0.829--0.832) \\
\bottomrule
\end{tabular}
\end{table}

\subsection*{Whole-night ECG-SDI trajectories reveal continuous variation within conventional sleep stages}

\begin{figure}[tbp]
\centering
\includegraphics[width=\textwidth]{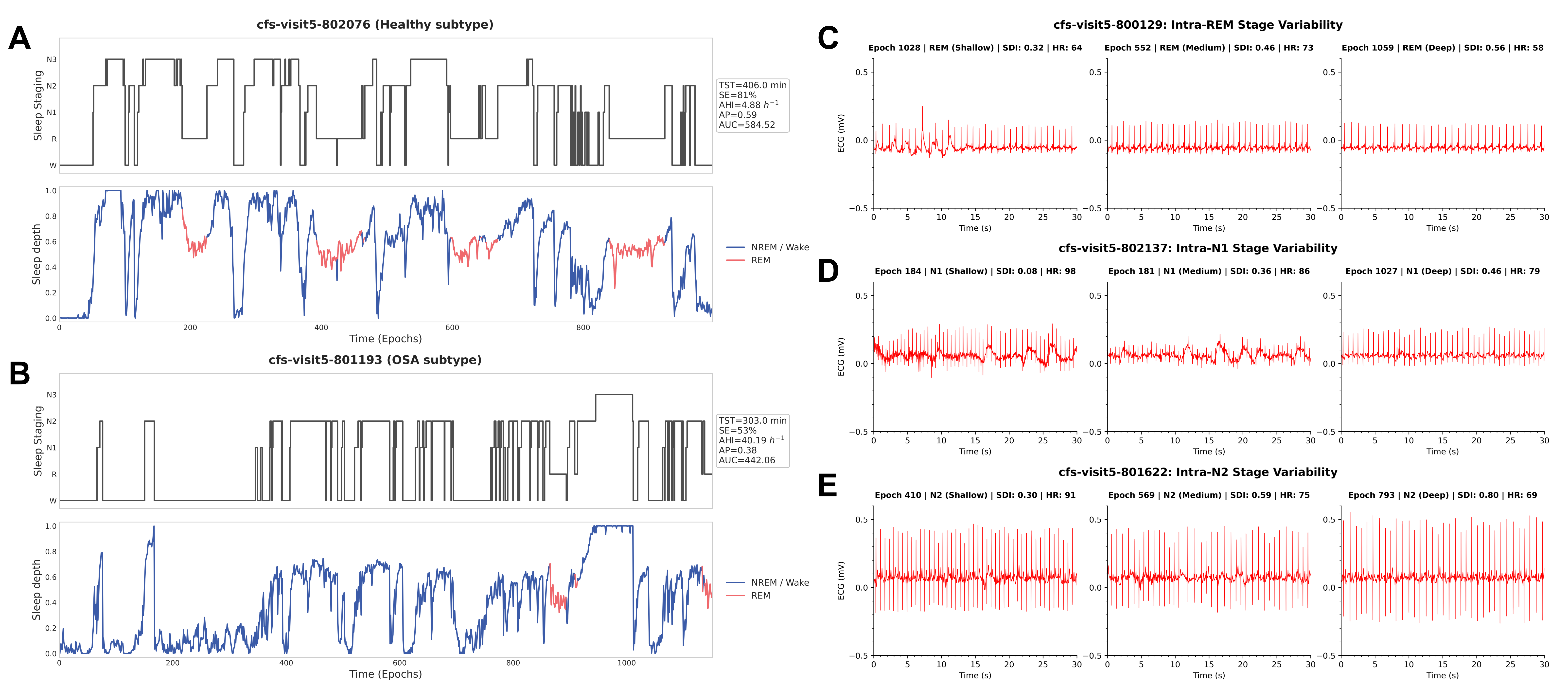}
\caption{\textbf{Whole-night ECG-SDI trajectories and within-stage ECG examples in the external CFS cohort.} \textbf{A,B,} Conventional hypnograms (upper panels) and corresponding ECG-SDI trajectories (lower panels) for a lower-AHI recording (\textbf{A}; cfs-visit5-802076) and an OSA recording (\textbf{B}; cfs-visit5-801193). REM epochs are highlighted in red, and wake/NREM epochs are shown in blue. Insets report total sleep time (TST), sleep efficiency (SE), apnea--hypopnea index (AHI), normalized area under the depth curve (AP), and area under the depth curve (AUC). \textbf{C--E,} Representative 30-s ECG segments from REM (\textbf{C}; cfs-visit5-800129), N1 (\textbf{D}; cfs-visit5-802137), and N2 (\textbf{E}; cfs-visit5-801622). Within each row, three epochs from the same participant and conventional sleep stage are arranged in ascending order of ECG-SDI. Each segment is annotated with its epoch identifier, stage, SDI value, and heart rate (HR; beats\,min$^{-1}$). The labels shallow, medium, and deep indicate relative positions on the ECG-SDI scale within the illustrated stage and do not represent additional conventional sleep stages.}
\label{fig:whole_night_cases}
\label{fig:intra_stage_cases}
\end{figure}

Representative recordings from the external CFS cohort illustrate how ECG-SDI describes both whole-night sleep organization and variation within conventional sleep stages (Fig.~\ref{fig:whole_night_cases}A,B). In the lower-AHI example, ECG-SDI repeatedly approached the upper end of its range during N3 sleep, occupied an intermediate range during REM, and decreased during wakefulness. The trajectory continued to fluctuate within intervals assigned the same stage label. This recording had a total sleep time of 406.0\,min, sleep efficiency of 81\%, and an apnea--hypopnea index (AHI) of 4.88\,h$^{-1}$ (Fig.~\ref{fig:whole_night_cases}A). The OSA example showed more sustained low-SDI intervals and recurrent interruptions of higher-SDI periods, together with shorter total sleep time (303.0\,min), lower sleep efficiency (53\%), and higher AHI (40.19\,h$^{-1}$; Fig.~\ref{fig:whole_night_cases}B). The corresponding AP values were 0.59 and 0.38, respectively.

To examine variation at the epoch level, we jointly displayed the ECG waveform, ECG-SDI value, and heart rate for selected 30-s epochs from REM, N1, and N2 sleep (Fig.~\ref{fig:whole_night_cases}C--E). Each stage-specific triplet comprised epochs from the same participant with the same conventional stage label. Within N1, ECG-SDI values of 0.08, 0.36, and 0.46 corresponded to heart rates of 98, 86, and 79\,beats\,min$^{-1}$, respectively. Within N2, ECG-SDI values of 0.30, 0.59, and 0.80 corresponded to heart rates of 91, 75, and 69\,beats\,min$^{-1}$. Thus, higher ECG-SDI values were accompanied by lower heart rates in these selected NREM examples. The REM examples had ECG-SDI values of 0.32, 0.46, and 0.56 and heart rates of 64, 73, and 58\,beats\,min$^{-1}$, respectively, showing that heart rate did not vary monotonically with ECG-SDI across all illustrated stages. Together, these examples relate the continuous model output to the underlying ECG and show cardiac variation within identical stage labels.

\subsection*{ECG-SDI-derived sleep phenotypes show distinct clinical profiles and an association with all-cause mortality}

\begin{figure}[p]
\centering
\includegraphics[width=0.80\textwidth]{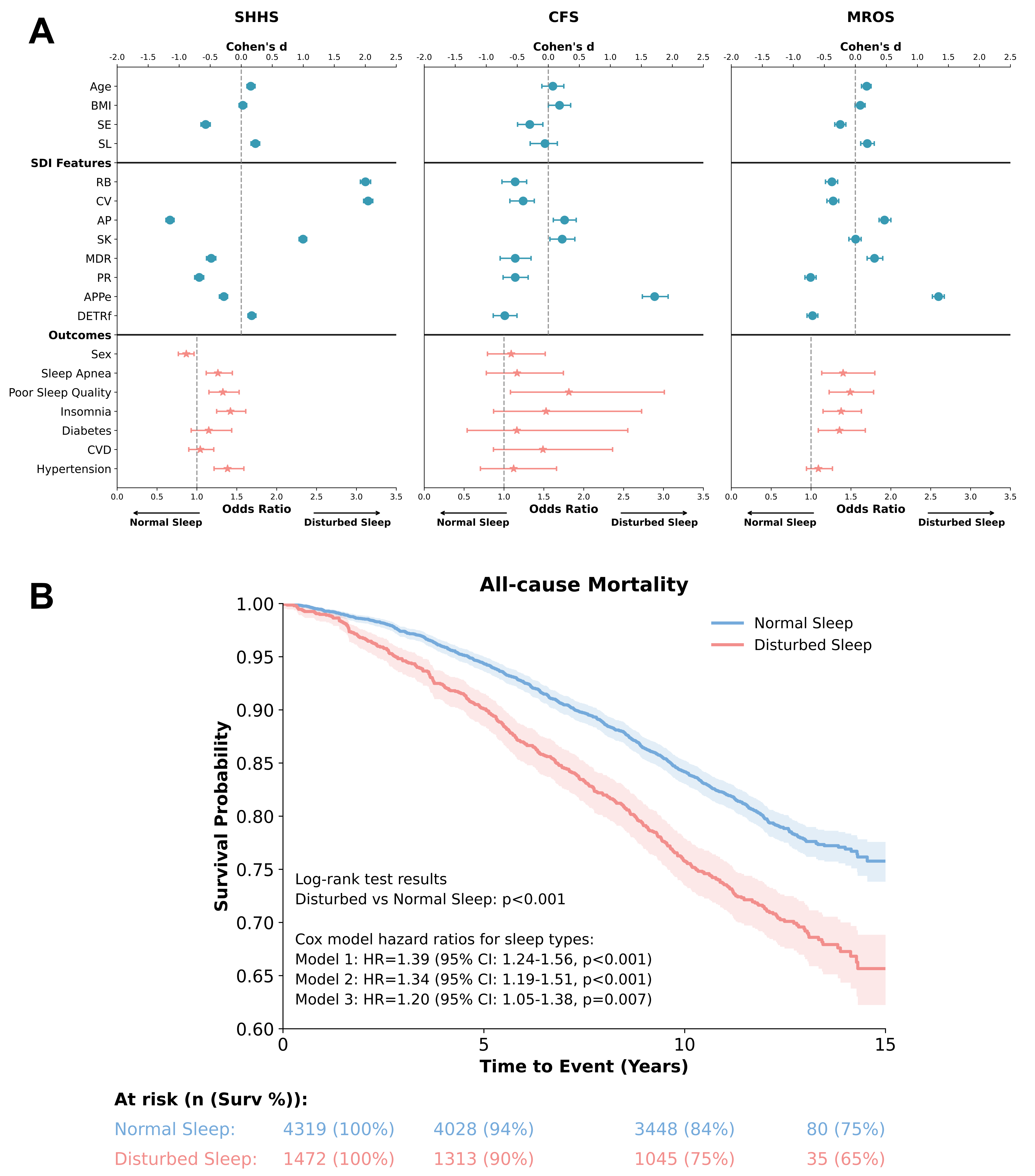}
\caption{\textbf{Clinical profiles and all-cause mortality associated with ECG-SDI-derived sleep phenotypes.} \textbf{A,} Comparisons between the normal sleep and disturbed sleep subtypes in SHHS, CFS, and MrOS. Upper panels show Cohen's \(d\) for demographic characteristics, conventional sleep measures, and ECG-SDI-derived biomarkers; positive values indicate higher values in the disturbed subtype. Lower panels show odds ratios for binary characteristics and clinical outcomes. Points and horizontal bars denote estimates and 95\% confidence intervals, respectively. Clinical outcome models were adjusted for age, body mass index, sex, sleep efficiency, and sleep latency, where applicable. Biomarker comparisons describe variables used for clustering. \textbf{B,} Kaplan--Meier curves for all-cause mortality in SHHS, with shaded 95\% confidence intervals and numbers at risk below. Blue and red indicate the normal sleep and disturbed sleep subtypes, respectively. The inset reports the log-rank test and Cox proportional hazards estimates. Model 1 adjusts for age, sex, and race; Model 2 additionally adjusts for body mass index, smoking status, hypertension, and diabetes mellitus; Model 3 additionally adjusts for apnea--hypopnea index and total sleep time.}
\label{fig:cluster_profiles}
\label{fig:survival_curves}
\end{figure}

Gaussian mixture modelling of whole-night ECG-SDI-derived biomarkers identified two sleep phenotypes in SHHS, CFS, and MrOS, designated the normal sleep and disturbed sleep subtypes. Their biomarker profiles differed across cohorts (Fig.~\ref{fig:cluster_profiles}A). In SHHS, the disturbed subtype had higher shallow-sleep burden and overnight variability (RB and CV; Cohen's \(d=2.00\) and \(2.05\), respectively) and lower cumulative sleep depth (AP; \(d=-1.15\)). In CFS, the largest separation was observed for approximate entropy (APPe; \(d=1.71\)); APPe was also higher in the disturbed subtype in MrOS. These profiles describe the cohort-specific combinations of ECG-SDI features underlying the two clusters.

The disturbed subtype had lower sleep efficiency across all three cohorts. After adjustment for age, body mass index, sex, sleep efficiency, and sleep latency, where applicable, this subtype was associated with poor subjective sleep quality in each cohort (Fig.~\ref{fig:cluster_profiles}A). The adjusted odds ratios were \(1.33\) (95\% confidence interval [CI], \(1.18\)--\(1.52\)) in SHHS and \(1.81\) (95\% CI, \(1.16\)--\(3.07\)) in CFS, with a positive association also observed in MrOS. Associations with insomnia symptoms were evident in SHHS and MrOS, while the disturbed subtype in SHHS also had greater odds of sleep apnea and hypertension. Thus, although the biomarker profiles varied across cohorts, the two phenotypes consistently differed in subjective sleep quality.

In the SHHS survival analysis, including 4,319 participants in the normal subtype and 1,472 in the disturbed subtype, Kaplan--Meier curves showed lower survival in the disturbed subtype (log-rank \(P<0.001\); Fig.~\ref{fig:cluster_profiles}B). The hazard ratio for all-cause mortality was \(1.39\) (95\% CI, \(1.24\)--\(1.56\)) in Model 1 and \(1.34\) (95\% CI, \(1.19\)--\(1.51\)) in Model 2 (both \(P<0.001\)). The association persisted in the fully adjusted model, which additionally included AHI and total sleep time (Model 3: hazard ratio, \(1.20\); 95\% CI, \(1.05\)--\(1.38\); \(P=0.007\)). These findings associate ECG-SDI-derived phenotypes with adverse sleep-health profiles and with long-term all-cause mortality in SHHS.

\subsection*{ECG-SDI biomarkers are treatment-responsive and track CPAP-associated restoration of sleep continuity}

\begin{figure}[h]
\centering
\includegraphics[width=0.95\textwidth]{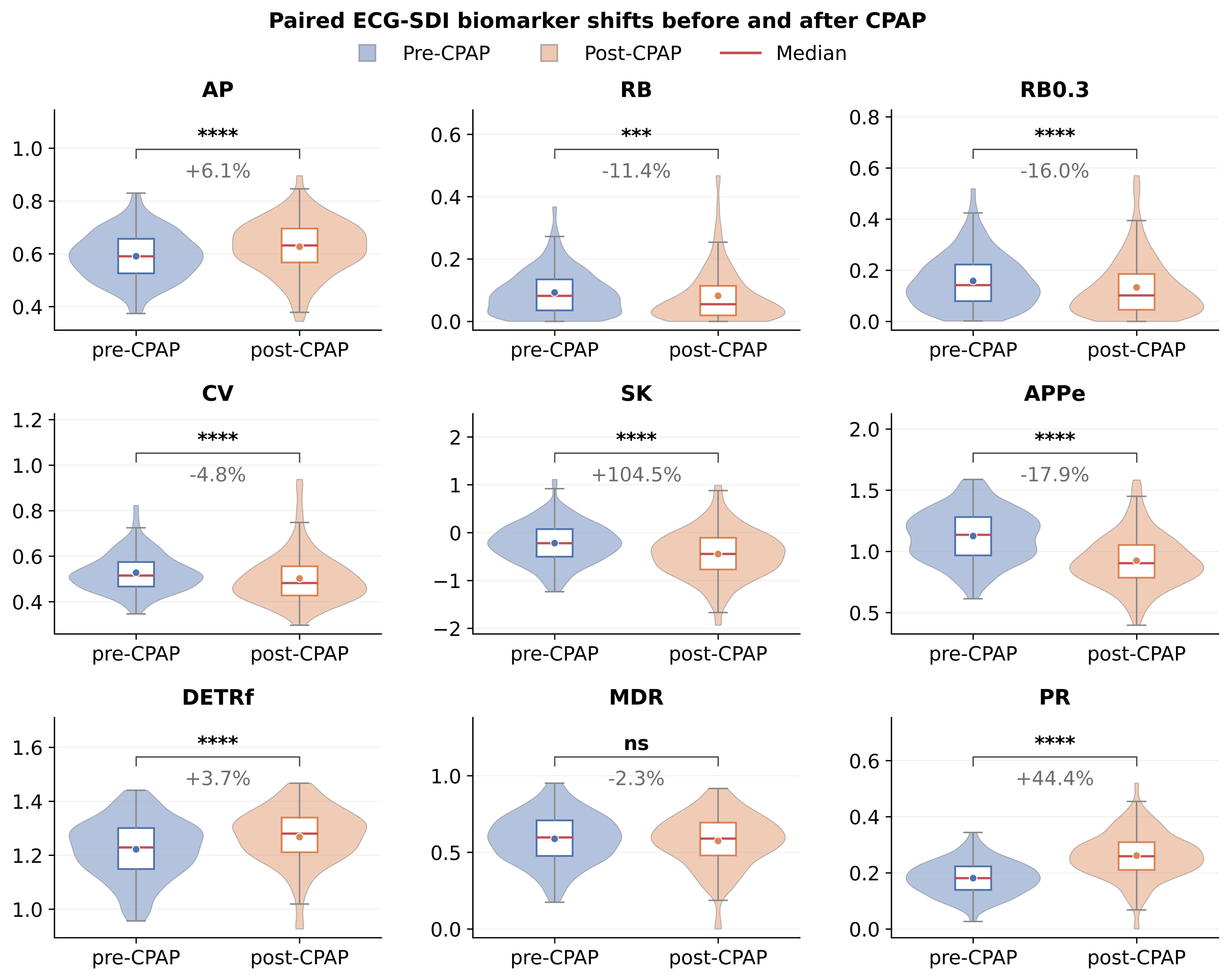}
\caption{\textbf{Paired distributions of key ECG-SDI-derived biomarkers before and after CPAP treatment in the BZ-CPAP cohort.}
Violin plots with embedded boxplots compare the distributions of AP, RB, RB\(_{0.3}\), CV, SK, APPe, DETRf, MDR, and PR between pre-CPAP and post-CPAP recordings. Red horizontal lines indicate medians, boxes denote interquartile ranges, and violin contours represent the full distribution of each biomarker. Across multiple metrics, post-CPAP values shifted toward higher accumulated sleep depth and reduced shallow-sleep burden, consistent with improved sleep continuity after treatment.}
\label{fig:cpap_violin}
\end{figure}

In the 507 paired pre- and post-treatment PSG recordings from the BZ-CPAP cohort, ECG-SDI-derived biomarkers showed a coherent post-CPAP shift consistent with improved overnight sleep continuity, which is shown in Fig.~\ref{fig:cpap_violin}. Compared with pre-CPAP values, AP, reflecting cumulative overnight sleep-depth burden, increased from 0.5909 to 0.6269, corresponding to a 6.1\% increase (\(p<0.001\)). In contrast, shallow-sleep burden decreased after treatment, with RB falling from 0.0926 to 0.0820 (\(-11.4\%\), \(p=0.0368\)) and RB\(_{0.3}\) declining from 0.1580 to 0.1327 (\(-16.0\%\), \(p<0.001\)). The overall variability of the overnight depth profile also decreased, as CV was reduced from 0.5277 to 0.5021 (\(-4.9\%\), \(p<0.001\)). Meanwhile, the proportion of REM sleep (PR) increased from 0.1811 to 0.2615 (\(+44.4\%\), \(p<0.001\)). DETRf also showed a modest increase, from 1.2219 to 1.2673 (\(+3.7\%\), \(p<0.001\)), whereas the change in MDR was not statistically significant (0.5880 vs.\ 0.5744, \(p=0.168\)). Detailed paired comparisons for the remaining complexity-related metrics are provided in Table~\ref{tab:appendix_cpap} of Appendix~\ref{app:cpap}. Taken together, these findings indicate that ECG-SDI-derived biomarkers shift after CPAP toward a sleep pattern that is deeper, less shallow-dominated, and less fragmented, supporting their utility as treatment-responsive digital biomarkers for tracking restoration of sleep continuity.

\subsection*{Exploratory analyses link CPAP-associated ECG-SDI changes to PSG improvement and within-stage variation}

\begin{figure}[p]
\centering
\includegraphics[width=\textwidth]{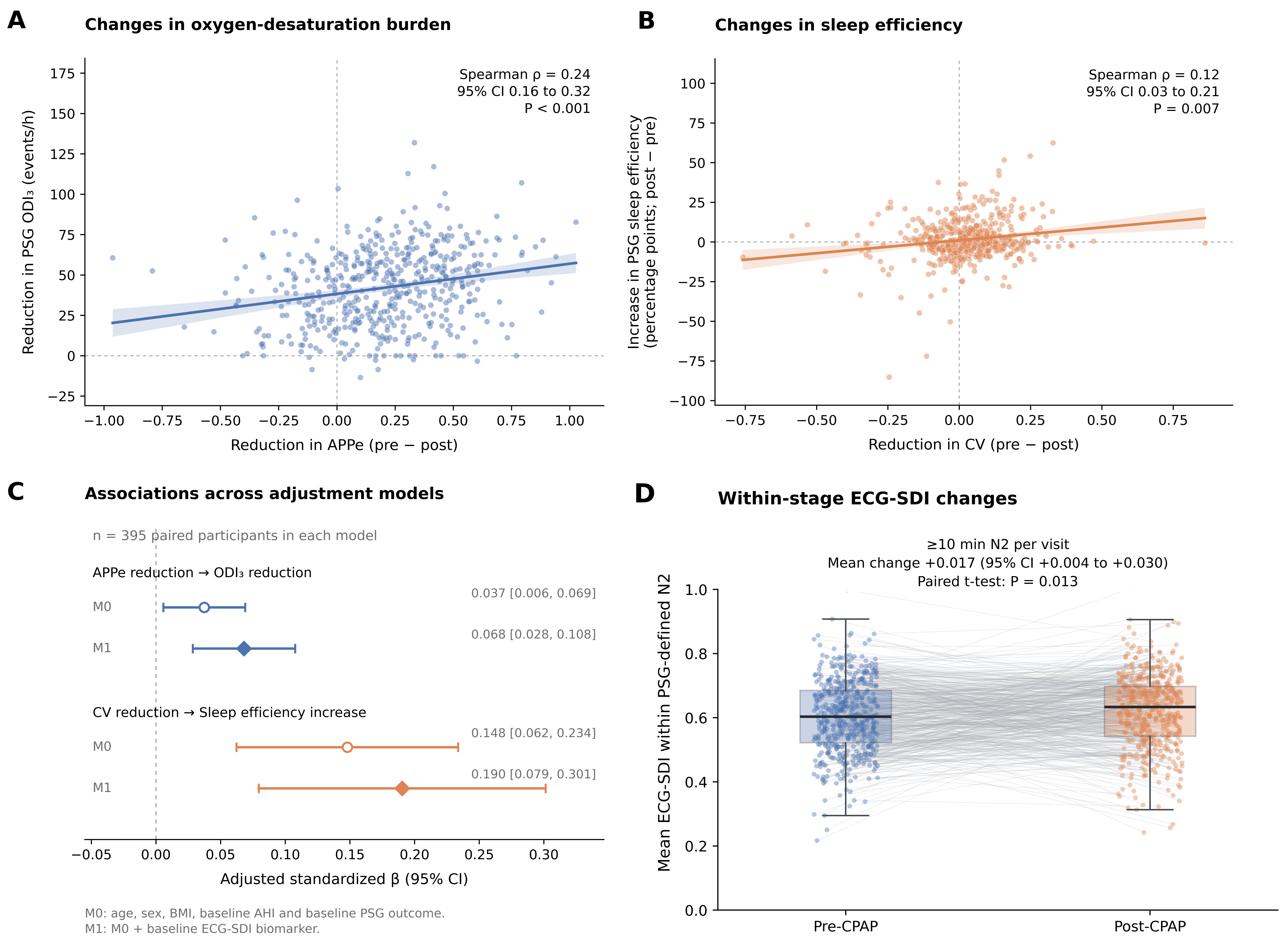}
\caption{\textbf{Exploratory associations of CPAP-associated ECG-SDI changes with PSG outcomes and within-stage sleep depth.} \textbf{A,} Association between reductions in APPe and PSG-derived ODI$_3$, both calculated as pre-CPAP minus post-CPAP values. \textbf{B,} Association between the reduction in CV (pre-CPAP minus post-CPAP) and the increase in PSG-derived sleep efficiency (post-CPAP minus pre-CPAP, in percentage points). Points represent paired participant-level changes; solid lines illustrate fitted trends. Spearman correlations, 95\% confidence intervals (CIs), and \(P\) values are reported in each panel. \textbf{C,} Adjusted standardized regression coefficients and 95\% CIs for the associations shown in A and B, with 395 paired participants included in each model. M0 adjusts for age, sex, BMI, baseline AHI, and the corresponding baseline PSG outcome; M1 additionally adjusts for the baseline value of the corresponding ECG-SDI biomarker. Open circles denote M0 and filled diamonds denote M1. \textbf{D,} Paired mean ECG-SDI values within PSG-defined N2 before and after CPAP, restricted to participants with at least 10\,min of N2 at each visit. Lines connect paired observations. The mean paired change, its 95\% CI, and the paired \(t\)-test \(P\) value are shown. APPe, approximate entropy; CV, coefficient of variation; ODI$_3$, oxygen desaturation index.}
\label{fig:cpap_exploration}
\end{figure}

We next examined whether CPAP-associated changes in ECG-SDI-derived biomarkers were related to changes in conventional PSG measures in the independent BZ-CPAP cohort. A greater reduction in approximate entropy (APPe) was associated with a greater reduction in the PSG-derived oxygen desaturation index (ODI$_3$; Spearman \(\rho=0.24\), 95\% CI, \(0.16\)--\(0.32\); \(P<0.001\); Fig.~\ref{fig:cpap_exploration}A). A greater reduction in the coefficient of variation (CV) was also associated with a greater increase in PSG-derived sleep efficiency, although the association was weaker (\(\rho=0.12\), 95\% CI, \(0.03\)--\(0.21\); \(P=0.007\); Fig.~\ref{fig:cpap_exploration}B). Reductions in biomarkers and ODI$_3$ were calculated as pre-CPAP minus post-CPAP values, whereas the increase in sleep efficiency was calculated as post-CPAP minus pre-CPAP values.

Both associations remained positive in covariate-adjusted analyses comprising 395 paired participants per model (Fig.~\ref{fig:cpap_exploration}C). After adjustment for age, sex, BMI, baseline AHI, and the corresponding baseline PSG outcome (M0), the standardized coefficients were \(\beta=0.037\) (95\% CI, \(0.006\)--\(0.069\)) for APPe reduction in relation to ODI$_3$ reduction and \(\beta=0.148\) (95\% CI, \(0.062\)--\(0.234\)) for CV reduction in relation to increased sleep efficiency. Additional adjustment for the baseline value of the corresponding ECG-SDI biomarker (M1) yielded coefficients of \(0.068\) (95\% CI, \(0.028\)--\(0.108\)) and \(0.190\) (95\% CI, \(0.079\)--\(0.301\)), respectively.

We additionally examined ECG-SDI within PSG-defined N2 among participants with at least 10\,min of N2 at each visit (Fig.~\ref{fig:cpap_exploration}D). Mean N2 ECG-SDI increased after CPAP by \(0.017\) (95\% CI, \(0.004\)--\(0.030\); paired \(t\)-test, \(P=0.013\)), indicating a small shift within a fixed conventional sleep stage. These exploratory analyses therefore identified modest associations between ECG-SDI-derived changes and PSG-defined improvement, together with a within-N2 increase in the model-derived sleep-depth index.

\subsection*{A wearable single-lead ECG workflow enables proof-of-concept home sleep-depth reporting}

\begin{figure}[tbp]
\centering
\includegraphics[width=\textwidth]{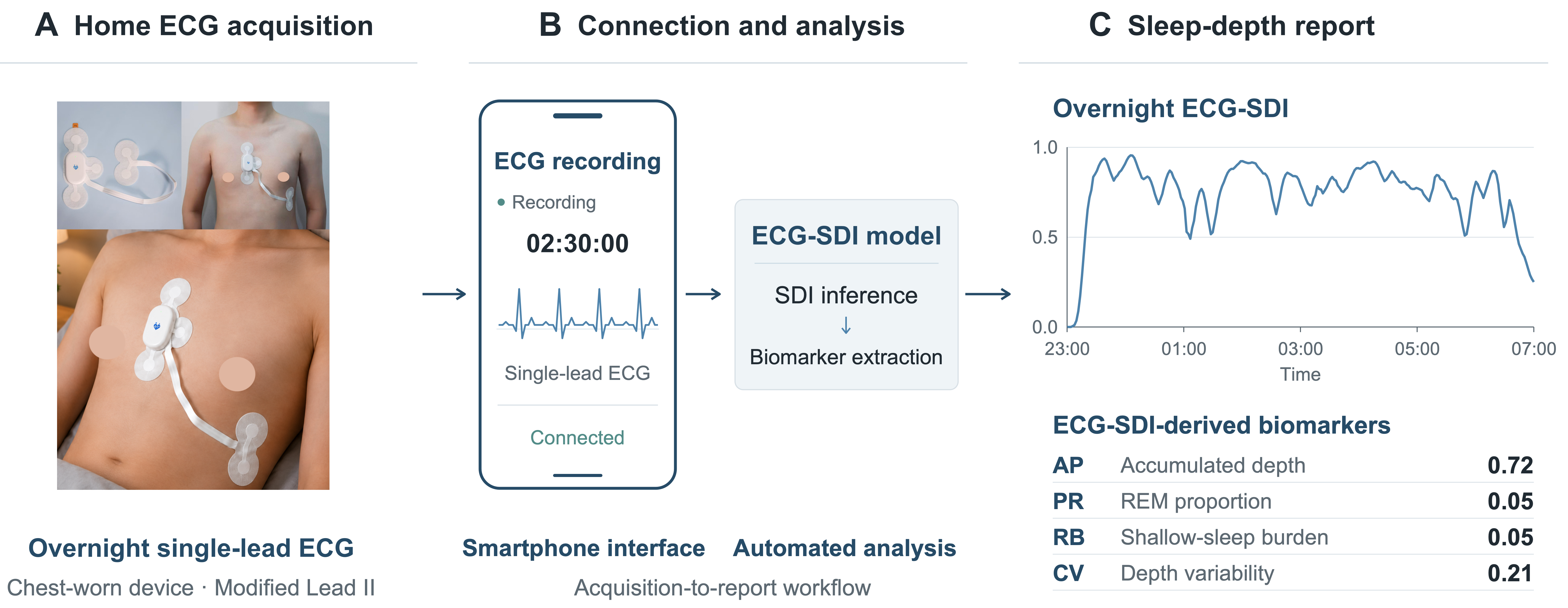}
\caption{\textbf{Proof-of-concept wearable workflow for home ECG-SDI reporting.} \textbf{A,} Illustration of overnight single-lead ECG acquisition using a chest-worn device in a modified Lead II configuration. \textbf{B,} Smartphone connection interface and automated analysis workflow comprising ECG-SDI inference and biomarker extraction. \textbf{C,} Illustrative report displaying an overnight ECG-SDI trajectory and summary values for AP, PR, RB, and CV. Interface elements, trajectories, and numerical values illustrate the workflow and do not constitute a home-validation dataset or validated normal/abnormal classifications.}
\label{fig:home_workflow}
\end{figure}

We implemented a proof-of-concept workflow integrating wearable single-lead ECG acquisition, automated ECG-SDI analysis, and sleep-depth reporting (Fig.~\ref{fig:home_workflow}). A chest-worn device in a modified Lead II configuration was linked to a smartphone interface for overnight recording (Fig.~\ref{fig:home_workflow}A,B). The acquired ECG was processed by the trained ECG-SDI model and biomarker-extraction module to generate an overnight depth trajectory and summary fields for AP, PR, RB, and CV (Fig.~\ref{fig:home_workflow}C). This implementation demonstrates the technical feasibility of an acquisition-to-report workflow for home monitoring. The displayed trajectory and biomarker values are illustrative; home-based estimation accuracy and individual treatment-monitoring performance were not evaluated.

% Complete revised Discussion. Uses existing reference keys plus qanash2017assessment and nelesen2001continuous.
\section*{DISCUSSION}

This study extends continuous sleep-depth phenotyping to single-lead ECG and evaluates its relevance to clinical sleep profiles, long-term outcomes, and treatment-associated changes. Across 12,392 overnight recordings, ECG-SDI recovered sleep-depth ordering and REM structure, characterized variation within conventional stages, and yielded biomarkers associated with subjective sleep quality and all-cause mortality. A central extension beyond our previous PSG-based work\cite{zhou2025continuous} is the evaluation of 507 participants before and after CPAP, providing evidence of treatment responsiveness alongside cross-sectional phenotyping. Together with a wearable acquisition-to-report proof of concept, these findings establish a basis for developing continuous sleep-depth assessment for repeated, lower-burden monitoring.

Continuous measures such as the EEG-derived odds ratio product and our previous SDI framework have shown that sleep depth contains information not fully represented by categorical stage labels.\cite{younes2015odds,zhou2025continuous} The present study extends this approach to a single cardiac signal, building on evidence that temporal patterns in cardiac activity support automated sleep staging.\cite{radha2019sleep} Correlations with ordered sleep stages (excluding REM) of 0.828 in the pooled internal test set and 0.848 in external CFS, together with REM AUROCs of 0.954 and 0.977, respectively, demonstrate that substantial sleep-related structure can be recovered without EEG input at inference. This reduction in sensor requirements is the main translational advantage: continuous phenotyping becomes compatible with an ECG recording modality that can be incorporated into lightweight wearable systems.

The whole-night and within-stage examples clarify what this continuous representation contributes. Overnight trajectories retain recognizable sleep architecture while displaying variation within periods assigned the same stage. Joint visualization of the ECG waveform, ECG-SDI, and heart rate makes these differences interpretable in relation to the recorded signal. Higher ECG-SDI accompanied lower heart rates in the selected N1 and N2 examples, whereas the REM examples showed no monotonic relationship. These observations are compatible with the stage-dependent cardiac autonomic patterns described in physiological studies,\cite{bonnet1997heart} while underscoring that ECG-SDI should not be interpreted as a simple inverse heart-rate scale. They illustrate variation within stage categories and motivate quantitative analyses of this heterogeneity.

The phenotype analyses extend this representation to clinically relevant differences between individuals. Despite differences in the biomarkers separating clusters across SHHS, CFS, and MrOS, the disturbed subtype consistently had poorer subjective sleep quality. Such heterogeneity is compatible with previous evidence that sleep-depth-related EEG biomarkers vary with demographic and clinical characteristics.\cite{younes2021characteristics} In SHHS, the association with all-cause mortality remained after adjustment including AHI and total sleep time (hazard ratio, \(1.20\); 95\% confidence interval, \(1.05\)--\(1.38\)). This finding supports the clinical relevance of the derived phenotype and motivates evaluation of whether continuous sleep-depth profiles add useful information to established sleep-health assessments. The cohort-specific biomarker patterns also favour interpreting these phenotypes as multidimensional profiles of sleep health.

The paired CPAP cohort provides a particularly valuable assessment of the responsiveness of ECG-SDI-derived biomarkers. CPAP is an established treatment for OSA,\cite{patil2019treatment} and previous EEG-based work has demonstrated changes in continuous sleep depth following treatment.\cite{penner2019odds} Here, comparing the same participants before and after CPAP reduced the influence of stable between-person differences and enabled evaluation of an ECG-based representation under clinically relevant treatment conditions. Increases in cumulative sleep depth and REM proportion, accompanied by reductions in shallow-sleep burden and overnight variability, formed a coherent pattern consistent with improved sleep continuity. The paired design thus extends the evidence for ECG-SDI from differences between individuals to changes within individuals, a necessary property for a candidate treatment-monitoring biomarker.

The exploratory analyses further relate these changes to conventional PSG outcomes. Reductions in APPe and CV were associated with improvements in ODI$_3$ and sleep efficiency, respectively, and the positive associations persisted after covariate adjustment. Their modest magnitudes (Spearman \(\rho=0.24\) and \(0.12\)) support physiological relevance without implying that these measures are interchangeable. The small increase in mean ECG-SDI within PSG-defined N2 (\(0.017\)) additionally indicates sensitivity to changes within a fixed stage category. This observation is relevant to prior ORP research showing that intervention-related changes within N2 contribute to overall NREM sleep-depth changes that stage durations alone incompletely describe.\cite{qanash2017assessment} CPAP can also alter cardiac autonomic regulation,\cite{nelesen2001continuous} so ECG-SDI changes may integrate sleep-related variation with accompanying cardiorespiratory responses. These findings support further evaluation of ECG-SDI alongside respiratory and conventional sleep measures when characterizing treatment response.

\section*{METHODS}

\subsection*{Study design, cohorts, and split strategy}

This was a retrospective multi-cohort study including 12,392 overnight PSG recordings. Among them, 10,652 recordings from the public SHHS\cite{quan1997sleep}, MESA\cite{chen2015racial}, and MrOS\cite{blackwell2011associations} cohorts available through the National Sleep Research Resource (NSRR)\cite{zhang2018national} were used for model development and were split at the participant level into training, validation, and internal test sets in a 60\%/20\%/20\% ratio. The CFS cohort\cite{redline1995familial} and the BZ-CPAP cohort contributed 1,740 PSG recordings for external evaluation and were not used for neural-network training or parameter selection. CFS contributed 726 recordings. Within BZ-CPAP, 507 participants underwent PSG before and after CPAP treatment, yielding 1,014 recordings for paired treatment-response analyses. The overall aim was to learn a continuous representation of sleep depth from single-lead ECG and to evaluate its stage ordering, clinical phenotyping value, and responsiveness to CPAP-associated changes. Cohort characteristics are provided in Appendix~\ref{app:info}.

\subsection*{ECG preprocessing and PSG-derived labels}

Single-lead ECG preprocessing followed the pipeline developed for our previous sleep-phenotyping framework\cite{xie2026holtertosleepaienabledrepurposingsinglelead}. In the public-cohort implementation, ECG signals were resampled to 100\,Hz and band-pass filtered at 0.5--40\,Hz. First-order polynomial detrending, robust standardization, and amplitude normalization were applied at the recording level, with the resulting signal scaled to \([-1,1]\). Signals were segmented into 30-s epochs and aligned with PSG sleep-stage annotations. Stages were encoded as Wake, N1, N2, N3, and REM; legacy stages 3 and 4 were combined as N3. In addition to the waveform, 39 frequency- and time--frequency-domain features were computed for each epoch. Both REM classification and depth ranking were supervised using discrete sleep-stage labels; manually annotated continuous depth targets were not required. Cohort-specific trimming, label handling, and feature construction are detailed in Appendix~\ref{app:repro_preprocessing}.

\subsection*{ECG-SDI framework and model development}

ECG-SDI used the sleep-staging backbone of our previously proposed two-stage single-lead ECG architecture\cite{xie2026holtertosleepaienabledrepurposingsinglelead}; the sleep-event detection branch was not used. In Stage 1, a Net1D-based convolutional encoder extracted time-domain representations from each 30-s ECG epoch\cite{hong2020holmes}. These representations were concatenated with an embedding of the 39 handcrafted features. In Stage 2, overlapping windows of 15 epoch embeddings were projected into tokens and processed as a sequence by a Transformer to incorporate temporal context\cite{vaswani2017attention}. Each token corresponded to the central epoch of its window. After sleep-staging pretraining, the convolutional feature extractor and Transformer weights were frozen, and the staging classifier was replaced by two trainable heads for REM classification and continuous depth estimation.

The joint objective adapted the ranking-based SDI framework established in our previous work\cite{zhou2025continuous}. REM cross-entropy was combined with a weighted pair-margin ranking loss over Wake and NREM stages. The depth head produced an unconstrained scalar. For trajectory visualization and biomarker extraction, depth scores were normalized within each recording using the 5th and 95th percentiles among reference Wake/NREM epochs and clipped to \([0,1]\). Thus, this research normalization used reference sleep-stage labels in addition to the ECG-derived model output. The unsmoothed normalized sequence was used for the reported public-cohort trajectory and biomarker analyses. Architecture, loss definitions, normalization, and implementation settings are specified in Appendix~\ref{app:repro_model}; the existing architecture diagram is provided in Fig.~\ref{fig:appendix_model_framework}.

\subsection*{ECG-SDI-derived biomarkers and phenotype discovery}

Following the digital biomarker framework established in our prior study\cite{zhou2025continuous}, each overnight ECG-SDI trajectory was summarized using time-domain, distributional, complexity, and REM-related descriptors. The coefficient of variation (CV) and skewness (SK) characterized the depth distribution. RB was the fraction of retained epochs with ECG-SDI below 0.2; RB$_{0.3}$ used a threshold of 0.3. AP was the trapezoidal area under the normalized ECG-SDI sequence divided by the number of retained epochs, including Wake. Approximate entropy (APPe) and detrended fluctuation analysis (DETRf) characterized temporal regularity and scaling behaviour\cite{richman2000physiological,hardstone2012detrended}. In the public-cohort implementation, mean REM depth (MDR) and REM proportion (PR) were calculated using reference PSG stage labels: MDR was the mean normalized depth during REM, and PR was the number of REM epochs divided by the number of non-Wake epochs. Mathematical definitions and computational conventions are provided in Appendix~\ref{app:repro_biomarkers}.

Eight biomarkers---AP, RB, CV, SK, APPe, DETRf, MDR, and PR---were standardized and entered into a two-component Gaussian mixture model separately within SHHS, CFS, and MrOS. After clustering, the component with higher mean AHI was designated the disturbed sleep subtype and the other component the normal sleep subtype. The two subtypes were compared in terms of biomarker profiles, clinical characteristics, and all-cause mortality. Clustering settings and clinical variable definitions are provided in Appendix~\ref{app:repro_statistics}.

\subsection*{Clinical definitions and statistical analysis}

Sleep apnea was defined as AHI \(\geq5\) events\,h$^{-1}$. Questionnaire-based sleep outcomes were derived from morning-after-PSG items. These included \texttt{diffa10}, \texttt{ltdp10}, and \texttt{rest10} in SHHS; \texttt{easlp}, \texttt{difbak}, \texttt{slpqua}, and \texttt{desslp} in CFS; and \texttt{poxfall}, \texttt{poxqual1}, and \texttt{poxqual3} in MrOS. The operational thresholds and combinations are specified in Appendix~\ref{app:repro_statistics}.

Model performance was evaluated using Spearman correlation between raw depth scores and ordered Wake/NREM stage labels, excluding REM, and the AUROC for REM classification. Public-cohort performance was calculated from pooled eligible epochs, with 95\% percentile confidence intervals estimated using 200 bootstrap resamples of epochs. Representative CFS recordings and within-stage examples were used for descriptive visualization. Within each stage-specific triplet, epochs were selected from the same participant and arranged by ECG-SDI. Heart rate was estimated from intervals between detected ECG peaks within the displayed 30-s segment, as described in Appendix~\ref{app:repro_preprocessing}.

Continuous subtype differences were summarized using Cohen's \(d\). Binary clinical associations were estimated using regularized logistic regression adjusted for age, BMI, and sex when available; sex was omitted when it was the outcome. Odds ratios were exponentiated subtype coefficients, and 95\% percentile confidence intervals were estimated using 500 bootstrap resamples. All-cause mortality in SHHS was assessed using Kaplan--Meier curves, log-rank tests, and Cox proportional hazards models based on NSRR follow-up records\cite{punjabi2009sleep}. Model 1 included age, sex, and race; Model 2 additionally included BMI, smoking, hypertension, and diabetes; Model 3 additionally included AHI and total sleep time. Estimation, coding, and censoring procedures are detailed in Appendix~\ref{app:repro_statistics}.

\subsection*{CPAP treatment-response and exploratory analyses}

Treatment-associated changes were evaluated using paired pre- and post-CPAP recordings from the BZ-CPAP cohort, with the participant as the analysis unit. Night-level biomarkers were compared using two-sided paired tests. Relative changes in the reported cohort means were expressed as \(100(\bar{x}_{\mathrm{post}}-\bar{x}_{\mathrm{pre}})/\bar{x}_{\mathrm{pre}}\); absolute changes were distinguished from percentage changes.

Exploratory analyses examined whether changes in ECG-SDI biomarkers were associated with changes in PSG outcomes. Reductions in APPe, CV, and ODI$_3$ were defined as pre-CPAP minus post-CPAP values, whereas increased sleep efficiency was defined as post-CPAP minus pre-CPAP values and expressed in percentage points. Spearman correlations were calculated for APPe reduction versus ODI$_3$ reduction and CV reduction versus increased sleep efficiency. Covariate-adjusted regression analyses included 395 paired participants per model. M0 adjusted for age, sex, BMI, baseline AHI, and the corresponding baseline PSG outcome; M1 additionally adjusted for the baseline value of the corresponding ECG-SDI biomarker. Associations were summarized by standardized coefficients and 95\% confidence intervals.

Within-stage changes were evaluated among participants with at least 10\,min of PSG-defined N2 at both visits. Mean ECG-SDI was calculated within N2 for each participant at each visit, and the paired post-minus-pre difference was assessed using a two-sided paired \(t\)-test and a 95\% confidence interval. Statistical significance was evaluated at \(P<0.05\), and the additional association and within-N2 analyses were interpreted as exploratory. The definitions and adjustment sets are consolidated in Appendix~\ref{app:repro_cpap}.

\subsection*{Wearable proof-of-concept implementation}
The proof-of-concept workflow linked chest-worn single-lead ECG acquisition in a modified Lead II configuration to a smartphone recording interface and ECG-SDI analysis and reporting components. ECG acquisition was performed using the Wen-Xin-Wu-Yang device (Anhui Heart Voice, China). The illustrated report comprised an overnight depth trajectory and summary fields for AP, PR, RB, and CV. This component demonstrated the proposed acquisition-to-report workflow; its displayed interface, trajectory, and numerical values were illustrative. No home-based accuracy analysis or individual treatment-benefit classification was performed. The PSG-referenced processing used for the research analyses is specified above and in the supplementary implementation details.

\newpage
%%%  The following components should appear after the 
%%%  methods. 
%%%  For journals using STAR Methods, these components 
%%%  should appear immediately after the discussion 
%%%  (after any "limitations" or "conclusions" subsection 
%%%  within the discussion).

\section*{RESOURCE AVAILABILITY}
\subsection*{Data availability}
The public sleep cohorts analyzed during the current study are available in the National Sleep Research Resource (NSRR) repository at \url{https://sleepdata.org/}. The BZ-CPAP dataset is not currently publicly available; inquiries regarding data access may be directed to the corresponding authors.

\subsection*{Code availability}
Code for public-cohort ECG preprocessing, sleep-staging backbone training, ECG-SDI adaptation and evaluation, biomarker extraction, phenotype clustering, survival analysis, and figure generation is available at \url{https://github.com/xdl1216/ecg-sdi}

\section*{ACKNOWLEDGMENTS}

This work was supported by the National Natural Science Foundation of China (62102008, 62172018), CCF-Tencent Rhino-Bird Open Research Fund (CCF-Tencent RAGR20250108), CCF-Zhipu Large Model Innovation Fund (CCF-Zhipu202414), PKU-OPPO Fund (BO202301, BO202503), Research Project of Peking University in the State Key Laboratory of Vascular Homeostasis and Remodeling (2025-SKLVHR-YCTS-02), Beijing Municipal Science and Technology Commission (Z251100000725008).

The Cleveland Family Study (CFS) was supported by grants from the National Institutes of Health (HL46380, M01 RR00080-39, T32-HL07567, RO1-46380). The National Sleep Research Resource was supported by the National Heart, Lung, and Blood Institute (R24 HL114473, 75N92019R002).

The Multi-Ethnic Study of Atherosclerosis (MESA) Sleep Ancillary study was funded by NIH-NHLBI Association of Sleep Disorders with Cardiovascular Health Across Ethnic Groups (RO1 HL098433). MESA is supported by NHLBI funded contracts HHSN268201500003I, N01-HC-95159, N01-HC-95160, N01-HC-95161, N01-HC-95162, N01-HC-95163, N01-HC-95164, N01-HC-95165, N01-HC-95166, N01-HC-95167, N01-HC-95168 and N01-HC-95169 from the National Heart, Lung, and Blood Institute, and by cooperative agreements UL1-TR-000040, UL1-TR-001079, and UL1-TR-001420 funded by NCATS. The National Sleep Research Resource was supported by the National Heart, Lung, and Blood Institute (R24 HL114473, 75N92019R002).

The National Heart, Lung, and Blood Institute provided funding for the ancillary MrOS Sleep Study, "Outcomes of Sleep Disorders in Older Men," under the following grant numbers: R01 HL071194, R01 HL070848, R01 HL070847, R01 HL070842, R01 HL070841, R01 HL070837, R01 HL070838, and R01 HL070839. The National Sleep Research Resource was supported by the National Heart, Lung, and Blood Institute (R24 HL114473, 75N92019R002).

The Sleep Heart Health Study (SHHS) was supported by National Heart, Lung, and Blood Institute cooperative agreements U01HL53916 (University of California, Davis), U01HL53931 (New York University), U01HL53934 (University of Minnesota), U01HL53937 and U01HL64360 (Johns Hopkins University), U01HL53938 (University of Arizona), U01HL53940 (University of Washington), U01HL53941 (Boston University), and U01HL63463 (Case Western Reserve University). The National Sleep Research Resource was supported by the National Heart, Lung, and Blood Institute (R24 HL114473, 75N92019R002).

\section*{AUTHOR CONTRIBUTIONS}

%%%  This component is required for most research papers. 
%%%  Mention each individual author with a statement 
%%%  outlining the contribution of each author to the work.

Donglin Xie conceived and designed the study, implemented the algorithms, analyzed the results, and drafted the manuscript. Qingshuo Zhao and Songchi Zhou contributed to data curation, preprocessing, and related supporting analyses. Haoyu Wang assisted with figure preparation and manuscript revision. Jingyu Wang provided the BZ-CPAP cohort data and offered clinical support. Yuxi Zhou contributed to methodological discussions and manuscript revision. Thomas Penzel and Shenda Hong supervised the project, provided overall technical guidance, and critically reviewed and revised the manuscript. All authors reviewed and approved the final manuscript.

\section*{DECLARATION OF INTERESTS}

%%%  This component is required for all articles, even 
%%%  if the authors have no competing interests; if 
%%%  this is the case, insert "The authors declare no 
%%%  competing interests." Please refer to the 
%%%  declaration of interests policy: 
%%%  https://www.cell.com/declaration-of-interests

The authors declare no competing interests.

\newpage

%%%  REFERENCES: As of 2023, all Cell Press journals 
%%%  use Numbered (AMA) style. We recommend placing 
%%%  your references in the included "references.bib" 
%%%  file.

\bibliography{reference}

\bigskip

%%%  In your References, please include only articles 
%%%  that are published (online publication and 
%%%  preprint servers are OK). Unpublished data, 
%%%  submitted and/or accepted manuscripts, abstracts, 
%%%  and personal communications should be cited within 
%%%  the text only ("unpublished data," "data not 
%%%  shown," "Alice Smith, personal communication") 
%%%  and not included in the references list. Personal 
%%%  communication should be documented by a letter 
%%%  of permission. Whenever possible, please make 
%%%  sure your .bib file has the complete author lists 
%%%  for each item (at minimum, the first 11 authors 
%%%  listed). 

\newpage
\begin{appendices}

\clearpage
\section{Baseline characteristics of the study cohorts}
\label{app:info}

Appendix Table~\ref{tab:appendix_cohort_info} summarizes the baseline characteristics of the cohorts included in this study, including sample size, age, female proportion, BMI, AHI, and acquisition timeframe. Except for female proportion, continuous variables are reported as mean \(\pm\) standard deviation; in the BZ-CPAP cohort, pre and post in the AHI column denote measurements obtained before and after CPAP treatment, respectively.

\begin{table}[h]
\centering
\caption{\textbf{Baseline characteristics of the study cohorts included in this work.} Continuous variables are reported as mean \(\pm\) standard deviation unless otherwise specified. In the BZ-CPAP cohort, pre and post in the AHI column denote measurements obtained before and after CPAP treatment, respectively.}
\label{tab:appendix_cohort_info}
\small
\setlength{\tabcolsep}{8pt}
\renewcommand{\arraystretch}{1.15}
\begin{tabularx}{\textwidth}{
l
>{\centering\arraybackslash}p{1.6cm}
>{\centering\arraybackslash}p{2.0cm}
>{\centering\arraybackslash}p{1.7cm}
>{\centering\arraybackslash}p{2.0cm}
>{\raggedright\arraybackslash}p{3.0cm}
>{\centering\arraybackslash}p{2.2cm}
}
\toprule
\textbf{Cohort} & \textbf{Number} & \textbf{Age (years)} & \textbf{Female (\%)} & \textbf{BMI (kg/m$^2$)} & \textbf{AHI} & \textbf{Timeframe} \\
\midrule
CFS      & 726   & 41.4 \(\pm\) 19.4 & 55\% & 32.4 \(\pm\) 9.5 & 10.2 \(\pm\) 18.5 & 2001--2006 \\
MrOS     & 2,806 & 76.4 \(\pm\) 5.5  & 0\%  & 27.2 \(\pm\) 3.8 & 13.7 \(\pm\) 14.4 & 2003--2012 \\
MESA     & 2,055 & 69.4 \(\pm\) 9.1  & 54\% & 28.7 \(\pm\) 5.5 & 14.8 \(\pm\) 16.7 & 2010--2013 \\
SHHS     & 5,791 & 63.1 \(\pm\) 11.2 & 52\% & 28.2 \(\pm\) 5.1 & 10.2 \(\pm\) 13.6 & 1995--2010 \\
BZ-CPAP  & 507   & 47.0 \(\pm\) 11.8 & 9\%  & 30.7 \(\pm\) 4.8 &
\makecell[l]{pre: 52.38 \(\pm\) 25.04 \\[2pt] post: 5.68 \(\pm\) 7.28}
& 2019--2025 \\
\bottomrule
\end{tabularx}
\end{table}

% =========================
% Appendix A: Complete table
% =========================

\clearpage
\section{Box plot  of stage-wise distribution of ECG-SDI}
\label{app:box}

\begin{figure}[h]
\centering
\includegraphics[width=\linewidth]{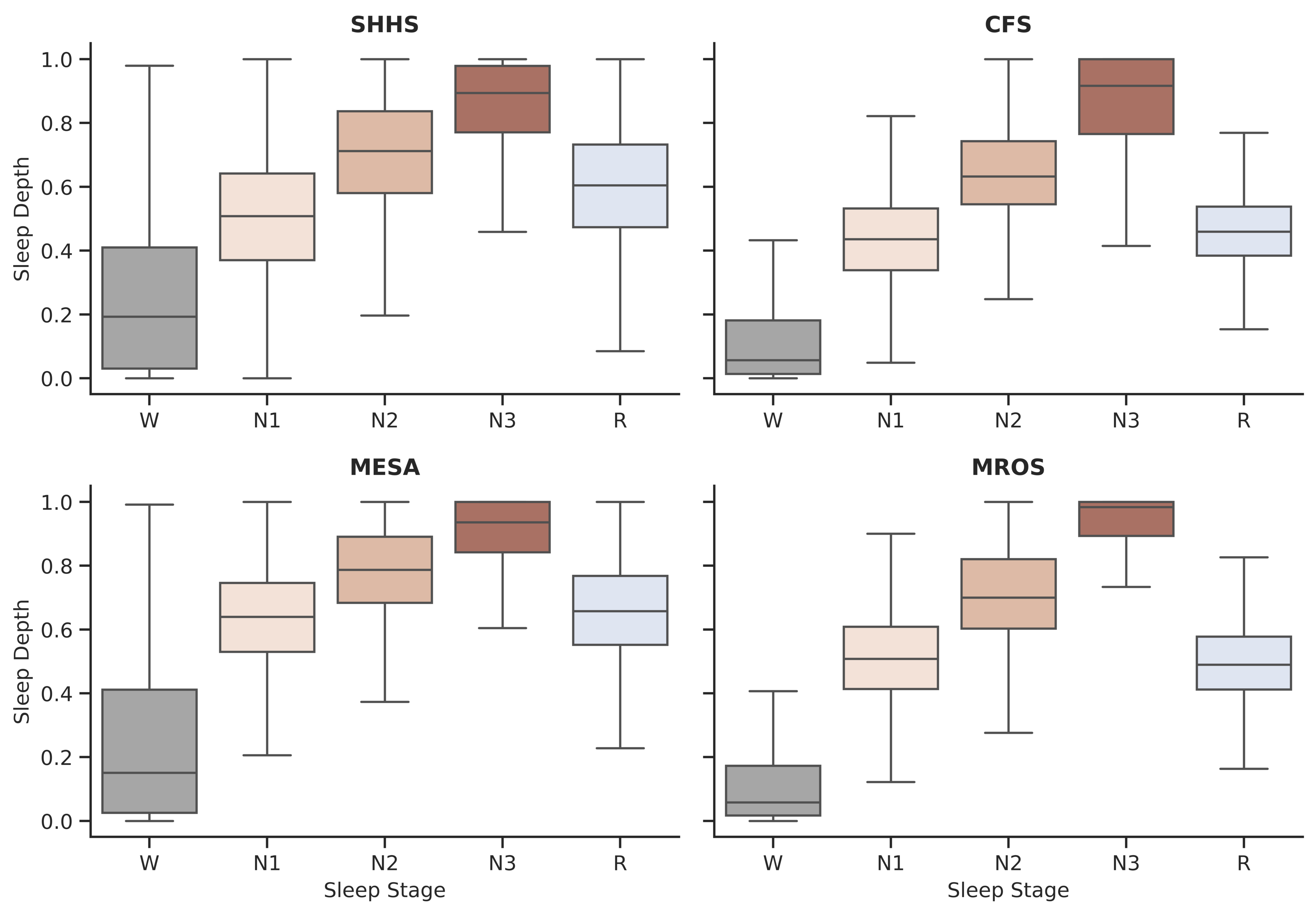}
\caption{Stage-wise distribution of ECG-SDI across conventional sleep stages in the SHHS, CFS, MESA, and MrOS cohorts. Boxplots show the distributions of ECG-SDI for wake (W), N1, N2, N3, and REM sleep. The center line denotes the median, the box denotes the interquartile range, and the whiskers indicate the distribution spread.}
\label{fig:stage_distribution}
\end{figure}

\clearpage

\begin{sidewaystable}[p]
\centering
\caption{\textbf{Detailed demographic, sleep, biomarker, and clinical characteristics of the two sleep phenotypes identified by clustering of sleep-depth-derived digital biomarkers.} For continuous variables, \textbf{N} and \textbf{D} denote mean \(\pm\) standard deviation values in the normal sleep subtype and disturbed sleep subtype, respectively, and \textbf{ES} denotes effect size with 95\% confidence intervals, calculated as Cohen’s \(d\). For binary clinical outcomes, the table reports the number of participants in each category for the two subtypes, and \textbf{ES} denotes adjusted odds ratios (ORs) with 95\% confidence intervals after controlling for age, BMI, sex, sleep efficiency, and sleep latency. In the sex row, the OR corresponds to female versus male; for the remaining clinical outcomes, ORs compare the prevalence of the outcome in the disturbed sleep subtype relative to the normal sleep subtype. MrOS was not included in the sex analysis because it is a male-only cohort, and CVD outcome data were not available in MrOS. This table provides the complete numerical counterpart of the effect-size and odds-ratio visualizations shown in Fig.~5 of the main text.}
\label{tab:appendix_cluster}
\setlength{\tabcolsep}{4pt}
\scriptsize
\resizebox{\linewidth}{!}{%
\begin{tabular}{@{}lcccccccccccc@{}}
\toprule
& \multicolumn{4}{c}{\textbf{SHHS}} & \multicolumn{4}{c}{\textbf{CFS}} & \multicolumn{4}{c}{\textbf{MrOS}} \\
\cmidrule(lr){2-5}\cmidrule(lr){6-9}\cmidrule(lr){10-13}
\textbf{Variable} & \textbf{N} & \textbf{D} & \textbf{ES (95\% CI)} & \textbf{\(P\)} & \textbf{N} & \textbf{D} & \textbf{ES (95\% CI)} & \textbf{\(P\)} & \textbf{N} & \textbf{D} & \textbf{ES (95\% CI)} & \textbf{\(P\)} \\
\midrule

\multicolumn{13}{@{}l}{\textbf{Demographics}} \\
\makecell[l]{Age\\(years)} 
& 62.70\(\pm\)10.99 & 64.43\(\pm\)11.81 & 0.15 (0.10, 0.21) & \(<0.001\)
& 40.99\(\pm\)19.01 & 42.42\(\pm\)20.45 & 0.07 (-0.08, 0.23) & 0.401
& 76.80\(\pm\)5.39 & 78.01\(\pm\)5.77 & 0.35 (0.26, 0.49) & \(<0.001\) \\

\makecell[l]{BMI\\(kg/m\(^2\))} 
& 28.13\(\pm\)4.95 & 28.26\(\pm\)5.49 & 0.03 (-0.03, 0.08) & 0.420
& 31.93\(\pm\)8.76 & 33.65\(\pm\)11.35 & 0.18 (0.01, 0.37) & 0.059
& 27.09\(\pm\)3.73 & 27.54\(\pm\)4.30 & 0.12 (-0.01, 0.22) & 0.043 \\

\makecell[l]{Sex\\(Female / Male)} 
& 2222 / 2097 & 809 / 663 & 1.15 (1.02, 1.30) & 0.021
& 297 / 239 & 102 / 88 & 0.93 (0.67, 1.30) & 0.744
& NA & NA & NA & NA \\

\addlinespace
\multicolumn{13}{@{}l}{\textbf{Traditional sleep metrics}} \\
\makecell[l]{Sleep efficiency\\(\%)} 
& 84.26\(\pm\)8.45 & 78.39\(\pm\)14.22 & -0.57 (-0.63, -0.51) & \(<0.001\)
& 79.65\(\pm\)10.98 & 75.82\(\pm\)17.06 & -0.30 (-0.46, -0.12) & 0.004
& 77.80\(\pm\)10.20 & 66.03\(\pm\)16.12 & -1.04 (-1.19, -0.95) & \(<0.001\) \\

\makecell[l]{Sleep latency\\(min)} 
& 12.90\(\pm\)17.70 & 17.55\(\pm\)26.54 & 0.23 (0.16, 0.30) & \(<0.001\)
& 20.77\(\pm\)28.27 & 18.94\(\pm\)47.77 & -0.05 (-0.29, 0.19) & 0.619
& 21.13\(\pm\)20.30 & 37.55\(\pm\)50.71 & 0.60 (0.40, 0.73) & \(<0.001\) \\

\addlinespace
\multicolumn{13}{@{}l}{\textbf{SDI-derived biomarkers}} \\
RB 
& 0.11\(\pm\)0.03 & 0.20\(\pm\)0.08 & 2.00 (1.93, 2.08) & \(<0.001\)
& 0.28\(\pm\)0.10 & 0.23\(\pm\)0.12 & -0.53 (-0.74, -0.33) & \(<0.001\)
& 0.30\(\pm\)0.10 & 0.35\(\pm\)0.16 & 0.51 (0.37, 0.67) & \(<0.001\) \\

CV 
& 0.44\(\pm\)0.05 & 0.58\(\pm\)0.10 & 2.05 (1.98, 2.12) & \(<0.001\)
& 0.68\(\pm\)0.13 & 0.62\(\pm\)0.16 & -0.40 (-0.62, -0.22) & \(<0.001\)
& 0.70\(\pm\)0.13 & 0.78\(\pm\)0.22 & 0.51 (0.36, 0.65) & \(<0.001\) \\

AP 
& 0.61\(\pm\)0.05 & 0.54\(\pm\)0.07 & -1.15 (-1.21, -1.08) & \(<0.001\)
& 0.47\(\pm\)0.06 & 0.49\(\pm\)0.09 & 0.26 (0.06, 0.45) & 0.009
& 0.45\(\pm\)0.06 & 0.44\(\pm\)0.10 & -0.21 (-0.33, -0.06) & \(<0.001\) \\

SK 
& -0.69\(\pm\)0.40 & -0.29\(\pm\)0.41 & 1.00 (0.95, 1.06) & \(<0.001\)
& -0.07\(\pm\)0.32 & 0.02\(\pm\)0.53 & 0.22 (0.02, 0.41) & 0.035
& 0.00\(\pm\)0.30 & 0.20\(\pm\)0.58 & 0.54 (0.38, 0.67) & \(<0.001\) \\

MDR 
& 0.61\(\pm\)0.12 & 0.53\(\pm\)0.27 & -0.48 (-0.57, -0.42) & \(<0.001\)
& 0.47\(\pm\)0.07 & 0.40\(\pm\)0.23 & -0.53 (-0.75, -0.32) & \(<0.001\)
& 0.47\(\pm\)0.07 & 0.53\(\pm\)0.21 & 0.50 (0.29, 0.73) & \(<0.001\) \\

PR 
& 0.21\(\pm\)0.06 & 0.16\(\pm\)0.09 & -0.68 (-0.76, -0.62) & \(<0.001\)
& 0.18\(\pm\)0.06 & 0.14\(\pm\)0.09 & -0.53 (-0.70, -0.33) & \(<0.001\)
& 0.20\(\pm\)0.06 & 0.14\(\pm\)0.08 & -0.94 (-1.06, -0.81) & \(<0.001\) \\

APPe 
& 0.78\(\pm\)0.15 & 0.73\(\pm\)0.20 & -0.28 (-0.34, -0.22) & \(<0.001\)
& 0.55\(\pm\)0.13 & 0.86\(\pm\)0.27 & 1.71 (1.53, 1.91) & \(<0.001\)
& 0.60\(\pm\)0.13 & 0.75\(\pm\)0.23 & 1.02 (0.88, 1.15) & \(<0.001\) \\

DETRf 
& 1.30\(\pm\)0.08 & 1.31\(\pm\)0.09 & 0.17 (0.11, 0.22) & \(<0.001\)
& 1.32\(\pm\)0.08 & 1.25\(\pm\)0.13 & -0.70 (-0.88, -0.54) & \(<0.001\)
& 1.33\(\pm\)0.07 & 1.28\(\pm\)0.10 & -0.81 (-0.94, -0.69) & \(<0.001\) \\

\addlinespace
\multicolumn{13}{@{}l}{\textbf{Clinical outcomes}} \\
\makecell[l]{Sleep apnea\\(No / Yes)} 
& 2156 / 2139 & 642 / 813 & 1.26 (1.15, 1.40) & \(<0.001\)
& 341 / 195 & 107 / 83 & 1.16 (0.70, 1.57) & 0.091
& 426 / 1960 & 59 / 361 & 1.22 (0.93, 1.97) & 0.067 \\

\makecell[l]{Poor sleep quality\\(No / Yes)} 
& 3486 / 809 & 1120 / 335 & 1.33 (1.18, 1.52) & \(<0.001\)
& 501 / 35 & 167 / 23 & 1.81 (1.16, 3.07) & 0.023
& 2054 / 332 & 295 / 125 & 2.71 (2.13, 3.27) & \(<0.001\) \\

\makecell[l]{Insomnia\\(No / Yes)} 
& 3100 / 1195 & 936 / 519 & 1.42 (1.29, 1.58) & \(<0.001\)
& 503 / 33 & 172 / 18 & 1.53 (0.95, 2.59) & 0.170
& 1960 / 426 & 277 / 143 & 2.29 (1.91, 2.73) & \(<0.001\) \\

\makecell[l]{Diabetes\\(No / Yes)} 
& 4010 / 285 & 1335 / 120 & 1.15 (0.94, 1.39) & 0.044
& 467 / 67 & 158 / 31 & 1.16 (0.61, 1.97) & 0.397
& 2084 / 302 & 353 / 67 & 1.22 (0.88, 1.61) & 0.078 \\

\makecell[l]{CVD\\(No / Yes)} 
& 3437 / 858 & 1119 / 336 & 1.04 (0.92, 1.20) & 0.013
& 480 / 56 & 157 / 33 & 1.49 (0.90, 2.26) & 0.018
& NA & NA & NA & NA \\

\makecell[l]{Hypertension\\(No / Yes)} 
& 2568 / 1727 & 728 / 727 & 1.38 (1.26, 1.60) & \(<0.001\)
& 363 / 173 & 117 / 73 & 1.12 (0.83, 1.61) & 0.147
& 1214 / 1172 & 199 / 221 & 1.07 (0.87, 1.24) & 0.204 \\

\bottomrule
\end{tabular}%
}
\end{sidewaystable}

\clearpage

\clearpage
\section{Detailed paired comparison of ECG-SDI-derived biomarkers before and after CPAP treatment}
\label{app:cpap}
To complement the main-text results on CPAP responsiveness, this table summarizes the detailed paired statistics of ECG-SDI-derived biomarkers in the 507 pre- and post-treatment PSG pairs from the BZ-CPAP cohort, including mean values before treatment, mean values after treatment, mean changes, significance levels, and statistical test results. Overall, AP, DETRf, and PR increased after treatment, whereas RB, RB$_{0.3}$, CV, SK and APPe decreased, indicating greater cumulative sleep-depth burden, reduced shallow-sleep dominance, attenuated overnight variability, and systematic shifts in complexity-related features. In contrast, MDR did not show a statistically significant pre--post change. This table provides the full numerical support for the distributional differences illustrated in Fig.~\ref{fig:cpap_violin} of the main text.

\begin{table}[h]
\centering
\caption{\textbf{Detailed paired comparison of ECG-SDI-derived biomarkers before and after CPAP treatment in the BZ-CPAP cohort.} pre-CPAP, before treatment; post-CPAP, after treatment; Change, post-CPAP minus pre-CPAP. $P$ values were obtained from two-sided paired tests. Sig indicates significance level: * $P<0.05$, ** $P<0.01$, *** $P<0.001$, ns not significant.}
\label{tab:appendix_cpap}
\setlength{\tabcolsep}{8pt}
\small
\begin{tabular}{lccccc}
\toprule
Metric & pre-CPAP & post-CPAP & Change & $P$ value & Sig \\
\midrule
AP         & 0.5909  & 0.6269  & +6.10\% & $<0.001$ & *** \\
RB         & 0.0926  & 0.0820  & -11.45\% & 0.0368   & *   \\
RB$_{0.3}$ & 0.1580  & 0.1327  & -16.01\% & $<0.001$   & *** \\
CV         & 0.5277  & 0.5021  & -4.85\% & $<0.001$   & *** \\
SK         & -0.2192 & -0.4483 & -104.5\% & $<0.001$ & *** \\
APPe       & 1.1257  & 0.9243  & -17.89\% & $<0.001$ & *** \\
% SampEn     & 0.9466  & 0.7487  & -20.91\% & $<0.001$ & *** \\
DETRf      & 1.2219  & 1.2673  & +3.72\% & $<0.001$ & *** \\
MDR        & 0.5880  & 0.5744  & -2.31\% & 0.1680   & ns  \\
PR         & 0.1811  & 0.2615  & +44.40\% & $<0.001$ & *** \\
\bottomrule
\end{tabular}
\end{table}

\clearpage
\section{Detailed architecture of the ECG-SDI framework}
\label{app:framework}
\begin{figure}[h]
    \centering
    \includegraphics[width=0.97\textwidth]{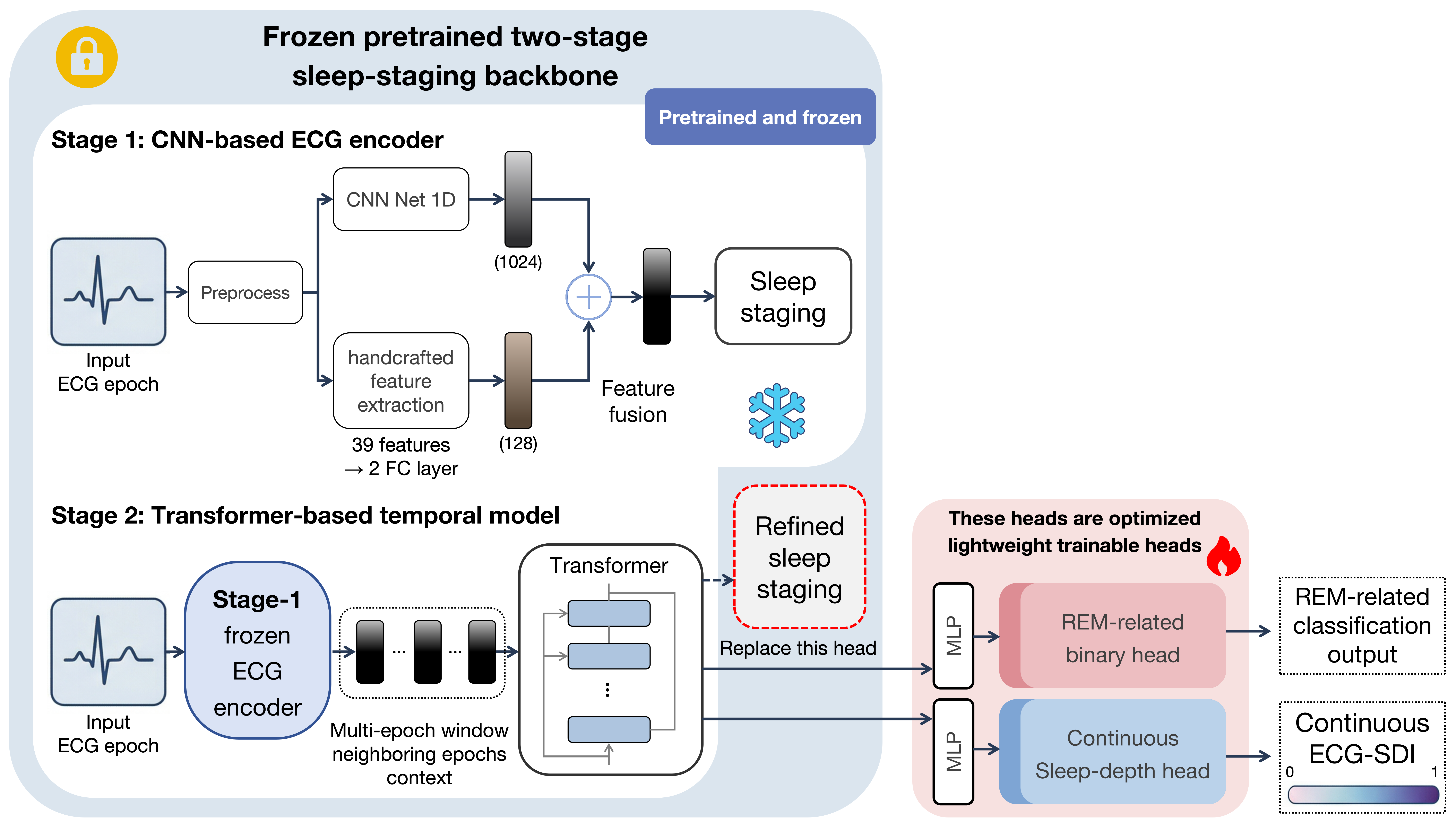}
    \caption{\textbf{Detailed architecture of the proposed ECG-SDI framework.} The proposed ECG-SDI framework is built upon a pretrained and frozen two-stage single-lead ECG sleep-staging backbone. In Stage 1, each 30-s ECG epoch is preprocessed and passed through a CNN-based time-domain encoder together with a handcrafted-feature branch, where 39 frequency- and time--frequency-domain features are projected through two fully connected layers and fused with the CNN representation to form an epoch-level sleep embedding. In Stage 2, the frozen Stage-1 encoder generates sequential multi-epoch representations, which are further processed by a Transformer-based temporal module to capture inter-epoch contextual dependencies and refine sleep representations. During ECG-SDI adaptation, the original refined sleep-staging head is replaced with two lightweight task-specific heads: one for REM-related binary classification and the other for continuous sleep-depth estimation, producing REM-related outputs and the continuous ECG-SDI score, respectively. This design repurposes a pretrained sleep-staging backbone for continuous sleep-depth biomarker learning while preserving multi-epoch temporal modeling.}
    \label{fig:appendix_model_framework}
\end{figure}

\clearpage
\section{Brief description of ECG-SDI-derived digital biomarkers}
\label{app:biomarkers}

Appendix Table~\ref{tab:appendix_biomarker_desc} summarizes the ECG-SDI-derived digital biomarkers used in this study together with brief interpretive descriptions. To improve readability, the descriptions emphasize the intuitive physiological meaning of each metric rather than its formal mathematical definition; among them, RB$_{0.3}$ and SampEn were primarily included as supplementary descriptors in the paired pre--post CPAP analyses.

\begin{table}[h]
\centering
\caption{\textbf{Brief description of ECG-SDI-derived digital biomarkers used in clustering and treatment-response analyses.} Descriptions are intended to provide intuitive physiological interpretation rather than formal mathematical definitions. RB$_{0.3}$ was additionally used in the CPAP treatment-response analysis.}
\label{tab:appendix_biomarker_desc}
\small
\setlength{\tabcolsep}{6pt}
\begin{tabularx}{\textwidth}{>{\raggedright\arraybackslash}p{1.8cm} >{\raggedright\arraybackslash}p{2.8cm} X}
\toprule
\textbf{Metric} & \textbf{Category} & \textbf{Brief description} \\
\midrule
AP & Depth burden & Normalized area under the whole-night ECG-SDI curve. Higher values indicate greater cumulative sleep-depth burden across the night. \\

RB & Shallow-sleep burden & Proportion of epochs with ECG-SDI values below 0.2, reflecting the overall dominance of shallow sleep. \\

RB$_{0.3}$ & Shallow-sleep burden & Proportion of epochs with ECG-SDI values below 0.3. This is a less stringent shallow-sleep index used mainly for treatment-response analysis. \\

CV & Variability & Coefficient of variation of the overnight ECG-SDI trajectory, describing the relative extent of night-long fluctuation. \\

SK & Distribution shape & Skewness of the ECG-SDI distribution, reflecting whether the depth profile is shifted toward shallower or deeper values. \\

APPe & Complexity & Approximate entropy, used to characterize the regularity and predictability of the ECG-SDI time series. Lower values indicate a more regular trajectory. \\

% SampEn & Complexity & Sample entropy, a complementary complexity measure that quantifies irregularity with reduced sensitivity to self-matches. \\

DETRf & Scale-free structure & Detrended fluctuation analysis index, describing scale-free self-affinity and long-range structure in the ECG-SDI trajectory. \\

MDR & REM organization & Mean ECG-SDI value during REM epochs, reflecting the average depth characteristics within REM sleep. \\

PR & REM organization & Proportion of REM sleep relative to total sleep time, reflecting REM-stage contribution to the whole-night sleep structure. \\
\bottomrule
\end{tabularx}
\end{table}

\newpage

\section{ECG preprocessing, handcrafted features, and illustrative cases}
\label{app:repro_preprocessing}

For the public-cohort EDF recordings, one ECG channel was extracted using the cohort-specific channel names: ECG for SHHS; ECG2, ECG1, or EKG for MESA; ECG2 or ECG1 for CFS; and ECGR, ECG R, or ECG for MrOS. Signals were resampled to 100\,Hz, band-pass filtered at 0.5--40\,Hz, and detrended by subtracting a first-order polynomial fitted over the recording. Standardization used the mean and population standard deviation of samples between the recording's 2.5th and 97.5th amplitude percentiles. The standardized signal was subsequently mapped to \([-1,1]\) using its recording-level minimum and maximum. Thus, the waveform supplied to the model and displayed in the within-stage examples had normalized amplitude rather than calibrated voltage.

The signal was divided into 30-s epochs containing 3,000 samples. Reference annotations were encoded as W, N1, N2, N3, and REM, with legacy stage 3 and stage 4 annotations combined as N3. Epochs without valid stage labels were excluded. The first and last 60 epochs were also excluded; the MrOS implementation applied this boundary exclusion only when more than 120 epochs were available. Feature extraction omitted recordings containing more than 1,680 retained epochs. A 15-epoch sliding context, advanced by one epoch, was centered on each prediction, leaving the first and last seven retained input epochs without predictions. Prediction indices therefore refer to the cropped, retained sequence; waveform displays were aligned to their corresponding preprocessed input epochs by adding the seven-epoch center offset.

Each epoch additionally contributed 39 handcrafted features. Nine spectral features were computed from the Welch power spectral density (PSD), using a segment length equal to the 3,000-sample epoch: integrated power over 0.5--4, 4--8, and 8--13\,Hz; the sum of PSD values; peak frequency; spectral centroid; spectral bandwidth; spectral flatness; and Shannon entropy of the normalized PSD. Thirty wavelet features were derived using a four-level Daubechies-4 decomposition. For each of the five coefficient arrays, the squared-coefficient sum, mean, standard deviation, variance, maximum, and minimum were calculated.

Whole-night examples used unsmoothed normalized SDI values together with reference hypnograms; REM highlighting was determined by the reference annotations. Displayed total sleep time and sleep efficiency were calculated from the retained prediction epochs, and the displayed area under the curve was obtained by trapezoidal integration at unit epoch spacing. Within-stage examples comprised manually selected 30-s epochs from the same participant and reference stage, arranged by increasing SDI. Heart rate was estimated from the displayed waveform using local-maximum detection with a minimum peak separation of 0.4\,s and a minimum prominence of 1.2 times the epoch's amplitude standard deviation. The reciprocal of the mean interpeak interval was multiplied by 60 and rounded to the nearest integer in beats\,min$^{-1}$; no estimate was returned when fewer than two peaks were detected. These examples were used for descriptive visualization rather than population-level inference.

\clearpage

\section{Reproducible model specification and evaluation procedures}
\label{app:repro_model}

\subsection{Epoch representations and temporal context}
The released implementation uses 30-s ECG epochs sampled at 100\,Hz. The CNN has seven stages with channel widths of 64, 160, 160, 400, 400, 1024, and 1024, comprising 2, 3, 3, 4, 4, 5, and 5 blocks, respectively; the kernel size is 16 and the downsampling stride is 2. Global average pooling, feature attention, and a fully connected transformation yield a 1024-dimensional waveform representation. The 39 handcrafted descriptors comprise nine spectral descriptors and 30 wavelet descriptors. Spectral descriptors include power in the 0.5--4, 4--8, and 8--13\,Hz bands, summed spectral power, peak frequency, spectral centroid, bandwidth, flatness, and entropy. Four-level Daubechies-4 decomposition provides five coefficient arrays, each summarized by energy, mean, standard deviation, variance, maximum, and minimum. A 39--64--128 multilayer perceptron with ReLU activations and a feature-attention module yields a 128-dimensional embedding, concatenated with the waveform representation to produce 1152 features per epoch.

Each temporal input comprises 15 successive retained epochs centered on the target epoch, with stride 1. The resulting $15\times1152$ feature window is flattened and projected to 512 dimensions. Learned positional embeddings and three Transformer encoder layers, each with eight attention heads, a feedforward width of 2048, and dropout of 0.1, model the sequence of windows within each recording. Thus, attention operates across the recording-level sequence, beyond the local 15-epoch input window. The positional-embedding capacity is 2048 windows. The first and last seven retained input epochs have no centered-window predictions. Prediction indices refer to the retained input sequence; correspondence with the original recording additionally requires accounting for preprocessing exclusions.

\subsection{Training stages and task-specific outputs}
The CNN is first trained for five-class sleep staging. Its fixed representations are then used to train the staging Transformer. For ECG-SDI adaptation, the staging classifier is removed, the pretrained temporal backbone is frozen, and separate REM and depth branches are optimized. Each branch applies layer normalization, a 512--2048--512 multilayer perceptron with a GELU activation, and a residual connection, followed by layer normalization and a final linear output. The REM branch produces two logits, whereas the depth branch produces one unrestricted real-valued score. REM probability is obtained by softmax, with a threshold of 0.5 for binary classification; the negative class includes wake and all NREM stages. No sigmoid transformation is applied to the depth output in the released evaluation pipeline.

Let $y_i\in\{0,1,2,3,4\}$ denote wake, N1, N2, N3, and REM, respectively, and let $d_i$ be the raw depth score. For $N$ valid epochs in a training batch, the implemented ranking objective is
\begin{equation}
\mathcal{L}_{\mathrm{rank}}=\frac{1}{\binom{N}{2}}
\sum_{i<j}\mathbf{1}(y_i<4,\,y_j<4,\,y_i\ne y_j)
\left[m_{y_i,y_j}-\operatorname{sgn}(y_i-y_j)(d_i-d_j)\right]_{+},
\end{equation}
where $[a]_{+}=\max(a,0)$. Margins for wake--N1, wake--N2, wake--N3, N1--N2, N1--N3, and N2--N3 are 1.0, 1.5, 3.0, 0.5, 2.0, and 1.5, respectively. All pairs involving REM are excluded from the numerator, including wake--REM pairs. Equal-stage and REM-containing pairs contribute zero but remain in the denominator. The ranking loss is zero when fewer than two valid epochs are available. The joint objective is
\begin{equation}
\mathcal{L}=\mathcal{L}_{\mathrm{REM}}+0.5\mathcal{L}_{\mathrm{rank}},
\end{equation}
where $\mathcal{L}_{\mathrm{REM}}$ is unweighted binary cross-entropy implemented through two-class softmax cross-entropy. Padding labels are ignored in both losses. Backbone parameters are frozen during adaptation, although Transformer dropout remains active in training mode. No padding attention mask is supplied to the Transformer.

\begin{table}[htbp]
\centering
\caption{Training settings in the released research scripts.}
\label{tab:repro_training}
\small
\begin{tabular}{llll}
\hline
Setting & CNN staging & Transformer staging & ECG-SDI heads \\
\hline
Optimizer & Adam & Adam & Adam \\
Initial learning rate & $10^{-3}$ & $10^{-4}$ & $10^{-3}$ \\
Weight decay & $10^{-5}$ & 0 & 0 \\
Maximum training epochs & 50 & 100 & 100 \\
Early-stopping patience & 5 & 5 & 5 \\
Batch size & 256 ECG epochs & 64 recordings\textsuperscript{a} & 1 recording \\
Random seed & $666+\mathrm{rank}$ & 42 & 42 \\
\hline
\end{tabular}

\vspace{2pt}
\begin{minipage}{0.96\textwidth}
\footnotesize
\textsuperscript{a}The Transformer source-script default is 64 recordings; the accompanying example JSON configuration sets this value to 1. CNN training additionally uses ReduceLROnPlateau with a reduction factor of 0.3 and patience of 10. Checkpoints are selected by minimum validation loss.
\end{minipage}
\end{table}

\subsection{Depth normalization and performance evaluation}
For each recording, $q_{5}$ and $q_{95}$ are the fifth and 95th percentiles of raw depth scores among reference wake/NREM epochs. Normalized ECG-SDI is
\begin{equation}
s_i=\min\!\left\{1,\max\!\left[0,\frac{d_i-q_{5}}{q_{95}-q_{5}}\right]\right\}.
\end{equation}
All epochs are used to estimate these percentiles if no reference wake/NREM epochs are present; when $q_{95}\leq q_{5}$, the implementation clips $d_i-q_{5}$ to $[0,1]$. This retrospective normalization uses reference PSG staging. The network itself receives ECG-derived features, but the released normalized-SDI evaluation is not a label-free deployment pipeline. Unsmoothed normalized scores are used by the stage-distribution, within-stage, and biomarker scripts and are the preferred input to the overnight plot. A separate five-point moving average is exported using same-length convolution with zero-padded boundaries.

Depth--stage Spearman correlation and pairwise concordance are calculated from pooled raw scores for reference wake/NREM epochs. REM metrics pool epoch-level probabilities across recordings. Percentile 95\% confidence intervals use 200 bootstrap resamples of pooled epochs with seed 42; they do not resample participants. The concordance routine retains same-stage pairs, counts concordant ordered pairs and exact joint ties, and evaluates all pairs when their number is at most 50,000. Otherwise, 50,000 index pairs are generated with seed 42 by first sampling the earlier index uniformly and then sampling a later index uniformly. CFS is evaluated separately using the same pipeline.

\clearpage

\section{Operational definitions of ECG-SDI biomarkers}
\label{app:repro_biomarkers}

Biomarkers were calculated from the unsmoothed, normalized depth sequence exported as \texttt{depth\_score\_norm}. Let \(s_i\) denote its value at retained prediction epoch \(i\), \(N\) the number of retained epochs, \(\bar{s}=N^{-1}\sum_i s_i\), and \(m_k=N^{-1}\sum_i(s_i-\bar{s})^k\). Unless otherwise specified, the calculations included all retained epochs, including wake and REM; no additional sleep-only restriction was applied. The operational definitions were
\begin{align}
\mathrm{AP} &= \frac{1}{N}\sum_{i=1}^{N-1}\frac{s_i+s_{i+1}}{2},\\
\mathrm{RB}_{q} &= \frac{1}{N}\sum_{i=1}^{N}\mathbb{I}(s_i<q),\qquad q\in\{0.2,0.3\},\\
\mathrm{CV} &= \frac{\sqrt{m_2}}{\bar{s}+10^{-6}},\qquad
\mathrm{SK}=\frac{m_3}{m_2^{3/2}}.
\end{align}
Here, \(\mathrm{RB}=\mathrm{RB}_{0.2}\), and \(\mathbb{I}(\cdot)\) is the indicator function. AP used trapezoidal integration with unit epoch spacing and normalization by \(N\), rather than by PSG total sleep time. SK followed the default moment-based estimator in \texttt{scipy.stats.skew}, without finite-sample bias correction.

Approximate entropy (APPe) was computed using \texttt{antropy.app\_entropy} (AntroPy 0.1.9), with embedding dimension \(m=2\), delay one epoch, Chebyshev distance, tolerance \(r=0.2\sqrt{m_2}\), and self-matches included. Specifically, \(\mathrm{APPe}=\Phi_2(r)-\Phi_3(r)\), where \(\Phi_m(r)\) is the mean natural logarithm of the proportion of length-\(m\) embedded vectors within distance \(r\) of each template. DETRf was computed using \texttt{antropy.detrended\_fluctuation}. The mean-centered sequence was cumulatively summed, partitioned into non-overlapping windows, and linearly detrended within each window. Integer window sizes increased approximately geometrically from four epochs to \(N/10\), with a factor of 1.2; incomplete final windows were discarded at each scale. DETRf was the slope of log root-mean-square fluctuation against log window size, excluding scales with zero fluctuation.

Let \(\mathcal{R}\) and \(\mathcal{S}\) denote retained epochs labelled REM and non-wake, respectively, by the reference PSG annotations. The REM summaries were
\begin{equation}
\mathrm{MDR}=\frac{1}{|\mathcal{R}|}\sum_{i\in\mathcal{R}}s_i,
\qquad \mathrm{PR}=\frac{|\mathcal{R}|}{|\mathcal{S}|}.
\end{equation}
These calculations used the exported reference \texttt{depth\_label}, rather than the predicted REM class. MDR was assigned zero when no REM epoch was present, and PR was assigned zero when no sleep epoch was present. Both were missing if reference staging was unavailable. Consequently, MDR combined ECG-derived depth with PSG-defined REM intervals, whereas PR was a reference-stage summary included alongside the depth-derived features.

\newpage

\section{Clinical coding, phenotype estimation, and survival analysis}
\label{app:repro_statistics}

The phenotype analyses linked biomarker records to cohort clinical tables by participant identifier. SHHS analyses used visit 1; CFS used visit 5; MrOS biomarkers were linked to the visit-1 clinical table. The input tables were SHHS baseline version 0.20.0, SHHS cardiovascular follow-up summary version 0.19.0 (with version 0.20.0 accepted as a fallback), CFS version 0.7.0, and MrOS version 0.6.0. Eight features---AP, RB, CV, SK, APPe, DETRf, MDR, and PR---were standardized to zero mean and unit population standard deviation separately within each cohort. Records missing any selected feature were excluded. A two-component Gaussian mixture model was fitted independently in each cohort using scikit-learn 1.7.2 defaults, including full covariance matrices, one k-means initialization, and random seed 42. The component count was fixed at two. After fitting, the component with higher mean AHI was labelled the disturbed sleep subtype and the other the normal sleep subtype; AHI was used for label orientation, not as a clustering input.

Questionnaire indicators were operational binary definitions of symptoms or perceived sleep quality, not clinical diagnoses. Their coding is summarized in Table~\ref{tab:repro_clinical_coding}. Sleep apnea was defined by AHI \(\geq5\,\mathrm{h}^{-1}\), using \texttt{ahi\_a0h4} in SHHS and CFS and \texttt{poahi3} in MrOS. Hypertension and diabetes used \texttt{htnderv\_s1} and \texttt{parrptdiab} in SHHS, \texttt{htn} and \texttt{diadiag} in CFS, and \texttt{mhbp} and \texttt{mhdiab} in MrOS. The CVD indicator used \texttt{any\_cvd} in SHHS, denoting CVD recorded since baseline, and \texttt{cvd} in CFS.

\begin{table}[htbp]
\centering
\caption{Operational definitions of questionnaire-derived binary outcomes.}
\label{tab:repro_clinical_coding}
\small
\setlength{\tabcolsep}{4pt}
\begin{tabularx}{\textwidth}{p{1.1cm}XX}
\toprule
Cohort & Poor sleep quality & Insomnia symptoms \\
\midrule
SHHS & \texttt{ltdp10} \(\leq2\) and \texttt{rest10} \(\leq2\) & \texttt{diffa10} \(\geq1\) \\
CFS & \texttt{slpqua} \(\geq5\) and \texttt{desslp} \(\geq4\) & \texttt{easlp} \(>4\) and \texttt{difbak} \(=1\) \\
MrOS & \texttt{poxqual1} \(\leq2\) and \texttt{poxqual3} \(\leq2\) & \texttt{poxfall} \(>40\) min \\
\bottomrule
\end{tabularx}
\end{table}

Continuous differences were expressed as Cohen's \(d\), calculated as disturbed-minus-normal mean difference divided by the pooled within-group sample standard deviation. Its 95\% confidence interval used the 2.5th and 97.5th percentiles of 500 bootstrap replicates, resampling records independently within each subtype. Binary associations used logistic regression with age, BMI, subtype, and sex when available; sex was omitted when it was the outcome. Models used an intercept and scikit-learn's default \(L_2\) penalty (\(C=1\)), with a maximum of 1,000 iterations. ORs were exponentiated subtype coefficients; percentile confidence intervals used 500 bootstrap resamples of the model-specific analytic records. Subtype assignments were held fixed during bootstrapping, and seed 42 was used. Missing continuous observations were omitted for each comparison. The implementation encoded missing binary clinical fields as zero and comparisons with missing questionnaire or AHI values as false; remaining missing covariates were excluded per regression. No multiplicity adjustment was applied to these intervals.

All-cause mortality in SHHS was defined by \texttt{vital}=0. Follow-up ran from the baseline PSG to death or last contact, using \texttt{censdate}/365.25 years. Participants without recorded death were censored at last contact. Kaplan--Meier estimates and two-sided log-rank tests compared subtypes. Cox models included age, sex, and numeric race code in model 1; BMI, hypertension, diabetes, and the available smoking variable were added in model 2; AHI and available total sleep time were added in model 3. Total sleep time was the sum of available \texttt{timest1}, \texttt{timest2}, \texttt{timest34}, and \texttt{timerem} fields. Smoking was selected as the first field whose name contained \texttt{smok}, \texttt{pack}, or \texttt{cigar} in the merged table. Missing hypertension, diabetes, AHI, race, and smoking values were coded zero; remaining missing model values were excluded, and constant covariates were removed. Cox fitting used \texttt{lifelines} 0.30.0 with an \(L_2\) penalizer of \(10^{-4}\); model-based 95\% confidence intervals and \(P\) values were obtained for the subtype coefficient.

\clearpage

\section{Definitions for exploratory CPAP analyses}
\label{app:repro_cpap}

\subsection{Participant-level changes and association models}
For participant $i$, reductions in each biomarker $X\in\{\mathrm{APPe},\mathrm{CV}\}$ and in ODI$_3$ were defined as
\[
\Delta X_i=X_{i,\mathrm{pre}}-X_{i,\mathrm{post}},\qquad
\Delta\mathrm{ODI}_{3,i}=\mathrm{ODI}_{3,i,\mathrm{pre}}-\mathrm{ODI}_{3,i,\mathrm{post}}.
\]
The increase in sleep efficiency was defined as
\[
\Delta\mathrm{SE}_i=\mathrm{SE}_{i,\mathrm{post}}-\mathrm{SE}_{i,\mathrm{pre}},
\]
with SE expressed as a percentage and its change therefore expressed in percentage points. Positive values indicated a reduction in APPe, CV, or ODI$_3$, or an increase in SE, respectively. The two reported change associations were $(\Delta\mathrm{APPe},\Delta\mathrm{ODI}_3)$ and $(\Delta\mathrm{CV},\Delta\mathrm{SE})$. Unadjusted associations were summarized by Spearman correlations and 95\% confidence intervals.

The adjusted analyses treated the change in the PSG measure as the outcome and the corresponding ECG-SDI biomarker change as the predictor of interest. For the APPe--ODI$_3$ association, M0 included age, sex, BMI, baseline AHI, and baseline ODI$_3$; M1 additionally included baseline APPe. For the CV--SE association, M0 included age, sex, BMI, baseline AHI, and baseline SE; M1 additionally included baseline CV. Each reported adjusted model included 395 paired participants. Regression associations were presented as standardized coefficients with 95\% confidence intervals. These analyses characterized associations between paired changes rather than individual treatment-response predictions.

\subsection{Mean depth within PSG-defined N2}
Let $\mathcal{N}_{i,v}$ denote the N2 epochs contributing ECG-SDI values for participant $i$ at visit $v\in\{\mathrm{pre},\mathrm{post}\}$. The participant-level N2 mean and paired change were
\[
\bar{s}_{i,v}^{\mathrm{N2}}=
\frac{1}{|\mathcal{N}_{i,v}|}\sum_{t\in\mathcal{N}_{i,v}}s_{i,v,t},
\qquad
\delta_i=\bar{s}_{i,\mathrm{post}}^{\mathrm{N2}}-\bar{s}_{i,\mathrm{pre}}^{\mathrm{N2}}.
\]
Eligibility required at least 10\,min of PSG-defined N2 at each visit. The analysis compared participant-level means using a two-sided paired $t$-test, with the mean paired difference and its 95\% confidence interval reported. Epochs were aggregated within each participant and visit before the paired comparison.

\end{appendices}

\end{document}